%% file: main.tex
\documentclass[twoside]{article}

\usepackage{PRIMEarxiv}

\usepackage{xcolor}
\usepackage{geometry} 
\usepackage{amsfonts}
\usepackage{amssymb}
\usepackage{cite}
\usepackage{graphicx}
\usepackage{caption}
\usepackage{subcaption}
\usepackage{algorithm}% http://ctan.org/pkg/algorithms
\usepackage{algpseudocode}% http://ctan.org/pkg/algorithmicx
\usepackage{tikz}
\usepackage{xparse}
\usepackage{wrapfig}
\usepackage[bottom]{footmisc}
\usepackage{bbm}
\usepackage{optidef}
\usepackage{mdframed}
\usepackage{enumerate}
\usepackage{hyperref}   

\usepackage{natbib}

\newtheorem{theorem}{Theorem}
\newtheorem{lemma}{Lemma}
\newtheorem{definition}{Definition}

\newtheorem{corollary}{Corollary}

\newcommand{\vq}{\mathbf{q}}
\newcommand{\vp}{\mathbf{p}}
\newcommand{\vv}{\mathbf{v}}
\newcommand{\vx}{\mathbf{x}}
\newcommand{\vX}{\mathbf{X}}

\newcommand{\vy}{\mathbf{y}}
\newcommand{\vY}{\mathbf{Y}}
\newcommand{\vz}{\mathbf{z}}
\newcommand{\vZ}{\mathbf{Z}}

\newcommand{\cl}{\mathcal}

\newcommand{\vlambda}{\boldsymbol{\lambda}}
\newcommand{\vrho}{\boldsymbol{\mu}}

\newcommand{\vone}{\boldsymbol{1}}
\newcommand{\vzero}{\boldsymbol{0}}

\newcommand{\bR}{\mathbb{R}}
\newcommand{\bZ}{\mathbb{Z}}

\newcommand{\cX}{\mathcal{X}}

\newcommand{\cD}{\mathcal{D}}
\newcommand{\cP}{\mathcal{P}}
\newcommand{\cS}{\mathcal{S}}

\newcommand{\cK}{\mathcal{K}}
\newcommand{\cT}{\mathcal{T}}
\newcommand{\cI}{\mathcal{I}}
\newcommand{\cJ}{\mathcal{J}}

\newcommand{\cU}{\mathcal{U}}
\newcommand{\cN}{\mathcal{N}}

\usepackage{comment}

\newcommand{\owner}[1]{}
\renewcommand{\owner}[1]{\colorbox{yellow}{\textbf{{#1}}}} 

\title{Middle-mile Optimization for Next-day Delivery}
\title{Middle-mile Truck Scheduling for Next-day Delivery Services}
\title{Middle Mile Truck Scheduling for Next Day Delivery}
\title{Last Truck Scheduling for Middle-mile Next-day Delivery Coverage}

\author{
	Konstantinos Benidis \\
	Amazon Research \\
	Berlin, Germany\\
	\texttt{kbenidis@amazon.com} \\
	\And
	Georgios Paschos\\
	Amazon Research \\
	Luxembourg\\
	\texttt{paschosg@amazon.lu} \\
	\And
	Martin Gross \\
	Amazon Research \\
	Luxembourg\\
	\texttt{grosmar@amazon.com} \\
	\And
	George Iosifidis\\
	Delft University of Technology \\
	Delft, Netherlands\\
	\texttt{g.iosifidis@tudelft.nl} \\
\thanks{Corresponding Author: G. Iosifidis. Address: Delft University of Technology, Mekelweg 5, 2628 CD Delft}	
}

\date{}

\begin{document}
	\maketitle

\begin{abstract}
We consider an e-commerce retailer operating a supply chain that consists of middle- and last-mile transportation, and study its ability to deliver products  stored in warehouses within a day from customer's order time. Successful next-day delivery requires inventory availability and timely truck schedules in the middle-mile and in this paper we assume a fixed inventory position and focus on optimizing the middle-mile last truck schedule. We formulate a novel optimization problem which decides the departure of the last truck at each (potential) network connection in order to maximize the number of customer orders that are served with next-day promise. We show that the respective \emph{next-day delivery optimization} is a combinatorial problem that is NP-hard to approximate within $(1-1/e)\cdot\texttt{opt}\approx 0.632\cdot\texttt{opt}$, hence every retailer that offers one-day deliveries has to deal with this complexity barrier. We study three variants of the problem motivated by operational constraints that different retailers encounter, and propose solutions schemes tailored to each problem's properties. To that end, we rely on greedy submodular maximization, pipage rounding techniques, and Lagrangian heuristics. The algorithms are scalable, offer worst-case optimality gap guarantees, and evaluated in realistic datasets and network scenarios were found to achieve even near-optimal results.

\end{abstract}

% Sample
%\KEYWORDS{deterministic inventory theory; infinite linear programming duality; 
%  existence of optimal policies; semi-Markov decision process; cyclic schedule}

% Fill in data. If unknown, outcomment the field

%------------------------------------------------------------------------------------------------------------------------------------------------------------------

%\tableofcontents

\input{intro-v3}

\input{related}
\input{problem-v3}

\input{complexity-v3}

\input{OB}

\input{IB}
\input{IOB}

\input{evaluation}

\input{discussion}

\vspace{5mm}   

\bibliographystyle{agsm}
\bibliography{deadline_opt.bib}

\newpage

\input{appendix}

\end{document}

%% file: intro-v3.tex
\section{Introduction}\label{sec:intro}

\subsection{Motivation}

Supply chains that support e-commerce like Amazon, Alibaba and Walmart are increasingly interested in optimizing their transportation network towards providing customers with improved next-day delivery (NDD) experience. A recent report by \citep{deshpande2022logistics} estimated the sales of Alibaba would increase by 13.3$\%$ if three-day deliveries (35$\%$ of orders) would be accelerated by one day. Apart from providing a competitive customer experience, increasing delivery speed was key for supporting economies during the COVID-19 pandemic  \citep{mollenkopf2020transformative}. Additionally, \citep{generation-report-2020} found that e-commerce produces 17\% less carbon emissions than customers visiting brick-and-mortar shops on a base scenario. Clearly, improving the range and volume of products that can be delivered within one day is poised to play a key role in shifting customer demand towards e-commerce, consolidating product supply and reducing carbon emissions. Therefore, it is not surprising that there is extensive work on same-day delivery (SDD) services \citep{savelsb-tutorial-TS16} where products are moved within a metro-area. Commonly encountered problems in this context are dynamic vehicle routing problems (VRP) \citep{klein2022dynamic}, optimization of delivery dispatches \citep{dispatch-TransScienc2018} and crowd-shipping \citep{carbone2017rise}, or the routing of (autonomous) vehicles \citep{otto2018optimization, figliozzi2020carbon}. In all these cases, the focus is on the last-mile transportation, whereas the middle-mile is less explored. For instance, \citep{konemann-time-expanded} studied the design of middle-mile network (selection of paths) for increasing delivery speed, while others have extended the classical VRP to hierarchical (e.g., 2-Echelon) networks \citep{Sluijk2023b}; see also Section \ref{sec:related-work}. Nevertheless, for the increasingly-encountered marketplaces with large selection of products that must be stored outside cities, the speed of delivery depends on different facets of the middle-mile network that extend beyond the selection of paths and hubs. 

This paper introduces a hitherto unexplored, yet crucial, middle-mile problem faced by such large e-retailers. We consider a general middle-mile scenario where upon customer order, middle-mile trucks transfer the products from origin warehouses (or Fulfillment Centers, FCs) to delivery stations (DSs), and last-mile vehicles deliver them to customers. These last-mile vehicles start the distribution at a given time of the day, hence a key task for the middle-mile is to inject the customer orders into the targeted DSs before these critical cutoff times. A natural objective here is to maximize the customer requests that are eligible for (are ``covered'' with) NDD, which is achieved whenever an order can be delivered by one of the middle-mile trucks to the respective DS on time. This presumes the availability of the product at some FC, and a truck departing from this FC \emph{after} the order is being processed and \emph{before} the last-mile cutoff at the targeted DS. In this context, our focus lies on the new problem of \emph{last truck dispatch} scheduling, i.e., to decide at which time slot each potential last truck for an l FC - DS connection should be scheduled (if any), towards maximizing the NDD demand coverage across all DSs.

The last departing trucks are obviously the last arriving trucks at the DSs, and in essence shape the NDD capabilities of the entire network.  Interestingly, the scheduling of last trucks affects also a number of other logistic processes, such as the profiling of labor capacity and design of labor shifts; the NDD ``promises'' offered to customers and their pricing; the inventory management at each FC and network connectivity, and so on.  In other words, in practice, the scheduling of last trucks are strategic decisions that inevitably have to be made at a coarse time scale, as they have direct implications for many key operations that, in turn, cannot be adjusted dynamically. On the other hand, the scheduling of the other (earlier) trucks can be decided in a much finer time scale, using  real-time information about the issued orders, volume of products, availability of truck and labor capacity, etc.; as well as taking into consideration the schedule of last trucks. At the same time, while decisions related to labor planning, inventory management, and network design are affected by the last truck dispatch, they are orthogonal as they operate in different time scales\footnote{E.g., the schedule of last trucks can be (and often is) re-optimized on a weekly basis, while consolidation paths are longer-term network configurations.} or simply provide input to this scheduling.

For this base scenario, we introduce and formulate an optimization problem, denoted $\cP$, that decides the last truck slots in order to maximize the customer orders that are served as NDD. The later these departures, the more customer orders are covered. While it is enticing to schedule these trucks exactly on the cutoff times, i.e., to leave from the FCs as late as possible, their schedule is subject to tight operational constraints, most notably the bounds on the number of trucks that can depart and/or arrive concurrently from/to the FCs/DSs\footnote{This bound in fact encompasses several operation limitations, e.g., the number of docking doors, the availability of labor, FC timing requirements, and so on.}. Similarly, more often than not, the FCs have overlapping inventory, and hence the demand at each DS can be covered by different and/or multiple FCs, which brings in a combinatorial aspect to the scheduling problem. Besides, in practice the middle-mile network is highly heterogeneous, as the FC - DS distances typically vary, which in turn shapes the last-slot availability for each potential connection. All these factors compound the last-truck scheduling decisions. In fact, we prove that this problem is NP-hard, and some of its variants are even inapproximable, and hence requires a tailored solution approach.

 \subsection{Contributions}
 
 In detail, we introduce the problem of maximizing the demand coverage of NDD services in the middle-mile by deciding the departure of last trucks (Section \ref{sec:problem-introduction}). We consider only direct FC - DS connections due to the need for expedite delivery and the hard deadlines that the last trucks, by definition, need to meet. We present three problem variants motivated by different operational constraints encountered in practice. Namely, we consider networks where: the Fulfillment Centers (origin nodes) have outbound capacity constraints, i.e., maximum number of concurrently-departing trucks; or the Delivery Stations (destination nodes) have inbound capacity constraints, namely maximum number of concurrently-arriving trucks; or both. We characterize the complexity of these problems and assess their properties in Section \ref{sec:theoretical}, where we prove that all variants are NP-hard to solve and, further, the inbound-constrained problem is hard to approximate to $\geq (1-1/e)\cdot \texttt{opt}$. In other words, designing the middle-mile last truck schedule so as to maximize coverage is a hard combinatorial problem, and this hinders its direct solution with off-the-shelf commercial MIP solvers when we encounter practical (thus, large) problem instances.

Having delved into their structure, we then propose a greedy and a pipage-based approximation algorithm, with performance guarantees of $1/2$ and $(1-1/e)$, respectively, which can solve the outbound-constrained (Section \ref{sec:ob}) and inbound-constrained (Section \ref{sec:ib}) variants. Moreover, we elaborate on their implementation details, introducing novel pipage execution strategies which trade-off runtime speed with performance (optimality gap). In Section \ref{sec:iob} we tackle the most compounded problem variant, i.e., the one with both inbound (at the DSs) and outbound (at the FCs) constraints, using Lagrangian heuristics. That is, we relax either the inbound or the outbound constraint, which creates a modified outbound (respectively, inbound) constrained problem with an objective that includes the relaxation penalties. Using a dual descent method with the Polyak step, we solve these problems to obtain solutions that are surprisingly-close to the solution found by commercial MIP solvers.
 
Finally, in Section \ref{sec:evaluation} we launch a series of experiments with carefully selected datasets and diverse network scenarios (see Appendix \ref{app:data}) in order to evaluate the proposed algorithms and compare them with meaningful benchmarks and commercial solvers. We find the performance of the algorithms to be even better than the above worst-case bounds while expediting the solution speed by orders of magnitude. We conclude and discuss future directions in Section \ref{sec:discussion}.

%% file: related.tex
\section{Literature Review}\label{sec:related-work}

Express shipping problems like SDD are typically classified as variants of VRPs with additional timing constraints. Unlike classical VRPs however, SDD models include dynamic arrival of orders, order cutoff time constraints, and delivery deadlines. Their goal is to maximize the orders served within each day, minimize penalties for undelivered orders, or minimize aggregate routing distance and service time. For example, \cite{dispatch-TransScienc2018} and \cite{dispatch-TransScienc2018b} schedule vehicle routes and dispatch waves in last mile networks under different assumptions. Similarly, \cite{ThomasBarrettW2019TSDP} optimize vehicle departures from a single depot in order to maximize the number of served requests. A different line of work focuses on tactical SDD decisions such as dimensioning the vehicle fleet, defining service deadlines or service areas, and so on, \cite{doi:10.1287/mnsc.2021.4041}, \cite{doi:10.1287/trsc.2022.1125}. On the other hand, the thrust of works about Two-Echelon VRPs go beyond the last mile by optimizing deliveries over an hierarchy of distribution centers, e.g., see \cite{Sluijk2023a} and references therein, including with timing constraints  \cite{dellaert2019}. Our work differs from these studies as we focus on truck scheduling in middle-mile with a demand coverage objective.

In relation to middle-mile, prior works focus on network design problems tailored to express shipment services. For instance, the hub-network design problem with time-definite delivery, decides hub locations and routing paths, where the deadlines are captured through path-eligibility constraints, see \cite{hub-location-campbell} and references therein. Similarly, \cite{hub-location-Transp21} study the NDD network design, and optimize the location of hubs, the assignment of demand centers to hubs, the routing and number of vehicles. Expanding on these ideas, \cite{wu-snd-ts23} consider additionally hub capacity constraints, such as restrictions on the concurrently loaded vehicles. Other variants of this problem include \citep{doi:10.1287/opre.1120.1065}, which optimizes truck release times to minimize the service cost while ensuring minimum NDD volumes; \citep{barnhart-express96}, which considers a single-hub overnight delivery system and optimizes routing subject to timing constraints; and \citep{yildiz-package-express}, which focuses on an express air service network design problem and decides which routes to operate with the company-owned cargo planes and how much capacity to purchase on commercial flights. All these works employ predetermined origin-destination (O-D) flows, and model the delivery time requirements through path-length constraints. Finally, \cite{dahan-lead-time} study the middle-mile consolidation problem with delivery deadlines while accounting for consolidation delays where, again, the O-D flows are predetermined. Similar models have been studied in the context of scheduled service network design that optimizes small (e.g., less-than-truckload) inter-city shipments with timing constraints for the delivery or intermediate hops, see \cite{hewitt-ssndp}.

% , since we are constrained by the operating schedule of these facilities limiting the demand that can be served within 1 day (next-day). Hence, unlike prior work, this paper's focus is not the trade-off between delivery speed and cost savings... \footnote{Put differently, for a given last truck schedule, the network can design a replenishment process to support the orders.}.

The goal of this paper is orthogonal to the above works. The O-D flows are not predetermined, nor we know the value of each flow, which, in fact, depends on the time-evolving demand and the FC(s) which serve that particular destination. This creates an intricate \emph{demand coverage problem}, which has not been studied before. Further, we optimize the \emph{last-truck} schedule using only the fast FC-DS direct paths (as opposed to slower multi-hop paths) as we are interested in coverage maximization, and we do not consider the trade-off between delivery speed and cost savings, that has been subject to several prior studies. Similarly, we do not consider inventory management and assume that a product's demand volume (before a cut-off time) at some DS can be covered by the selected FC if a last truck is scheduled Similarly, our goal is not devising a full truck schedule (i.e., across the entire day) but instead to optimize the truck cut-off times as these are the ones shaping the demand that can be covered, and because of the different time scales involved in these two schedules (last versus earlier trucks).

Finally, regarding our solution approach, we draw results and ideas from the theory of submodular optimization \citep{krause2014submodular} and employ the greedy algorithm, along with tailored pipage rounding steps \citep{ageev2004pipage}, to solve the problems at hand with optimality guarantees. Going a step further, we use Lagrange heuristics \citep{fischer-lagrange} for the most compounded problem version which, despite not offering such guarantees, are found to perform nearly-optimal (with respect to MIP solutions), with low overheads per iteration, when using the step proposed by \cite{polyak-step}. These solution speed gains have tremendous practical importance since, as we observe in the experiments, the solution of even medium-size instances requires multi-hours calls of commercial solvers in well-provisioned computing infrastructure. 

%% file: problem-v3.tex
\section{Problem Definition}\label{sec:problem-introduction}

In this section we introduce our supply chain model and state formally the next-day delivery (NDD) last truck optimization problem. To streamline the presentation we denote sets with uppercase calligraphic letters, i.e., $\cS$, and their cardinality as $S=|\cS|$. For indexing the set we use its lowercase letter, i.e., $s\in\cS$. Matrices are denoted with uppercase bold letters, vectors with lowercase bold letters and scalars with lowercase normal letters, i.e., $\vX$, $\vx$ and $x$, respectively. We use $\mathbf 1_T$ ($\mathbf 0_T$) to denote a vector of ones (zeros) of size $T$. 

\subsection{System Model}

\paragraph{Network:} The transportation network consists of a set $\cl I$ of $I=|\cl I |$ Fulfillment Centers (FCs) and a set $\cl J$ of $J=|\cl J|$ Delivery Stations (DSs). The distances among the FCs and DSs can vary;? and $\delta_{ij} \in \bR_+$ denotes the transit time or delay, i.e., the time a truck needs to travel, from FC $i\in \mathcal I$ to DS $j\in \mathcal J$. 
 
\paragraph{Products:} We denote by $\cK$ the set of $K=|\cK|$ available products and by $\cl K_i \subseteq \cK$ the subset of products stored in each FC $i\in \mathcal I$; such inventory limitations arise naturally due to space and storage constraints. We use the binary parameter $a_{ik}$ to denote whether product $k\in \cl K$ is stored at FC $i\in \mathcal I$, where $a_{ik} = 1$ means $k \in \cl K_i$. Based on the scenario and the e-retailer, the granularity of items can be defined at the product group level, or product category, i.e., $\cK$ may represent a set of product categories. Our analysis is oblivious to the definition of product granularity. Furthermore, we assume that a replenishment process runs in the background, such that if a product is in the inventory of an FC, it can indeed serve the demand at its assigned DSs. In other words, there is a replenishment strategy, orthogonal to this model, such that we do not experience periods of product unavailability\footnote{Otherwise, one can simple set $a_{ik}=0$ to denote coverage limitations for the decided plan.}.

\paragraph{Demand:} For a given time granularity (e.g., hourly timeslots) we define the corresponding set of timeslots $\cl T$, with $T=|\cl T|$. For each timeslot $t\in\cT$ we denote with $d_{jkt} \in \bZ_+$ the estimated new demand of product $k\in \cl K$ at DS $j\in \cl J$. This amounts to the additional demand (after slot $t-1$) that has been submitted by a customer in the area of DS $j\in\cJ$ and is ready to be shipped from slot $t\in\cT$ and onward from an FC.\footnote{In practice, this quantity corresponds to the new, i.e., after slot $t-1$, customer requests that have been processed and can be delivered by the selected FCs. In other words, the reference slot $t$ in $d_{jkt}$ does not necessarily reflect when the order are placed, but instead when they are ready to be processed by the middle mile.} 

\begin{figure}[t]
\centering
  \includegraphics[width=0.67\linewidth]{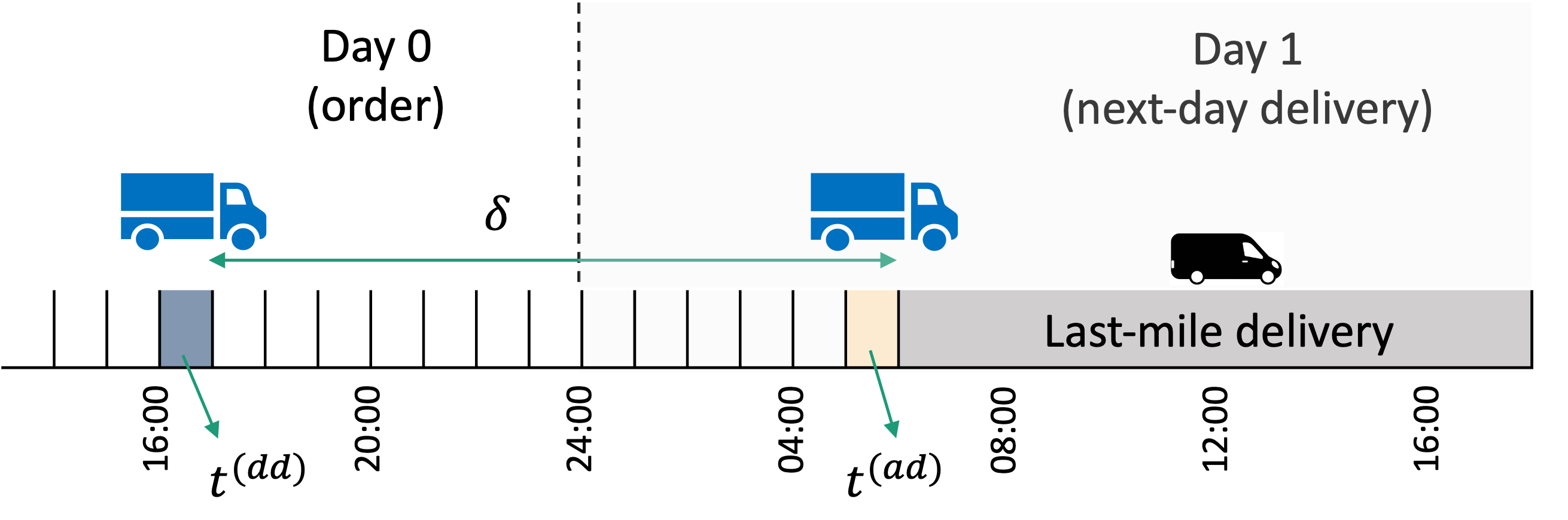}
  \caption{Example of a single FC-DS connection. The cutoff time of the DS (for NDD) is $t^\text{(ad)}=30$ (or next day 6am), the transit time $ \delta = 13$ and the deadline departure $t^\text{(dd)}=17$ (5pm).}
  \label{fig:last_truck}
\end{figure}

\paragraph{Arrival and departure deadlines:} We will study all DS injections that correspond to a  day of service in the last-mile. Therefore, $\cT$ spans all timeslots relevant to middle-mile departures and last-mile arrivals, as explained next. At each DS $j\in\cJ$ there exists a daily arrival deadline $t^\text{(ad)}_j$ which represents the critical timeslot for truck arrivals that can be processed on time to deliver their loads to customers with next-day promise; see Figure \ref{fig:last_truck} for a representation of how the last-mile cycle affects the middle-mile injections. Any trucks arriving after this cut-off $t^\text{(ad)}_j$, will not contribute to the NDD objective. We choose $\cT$ in a way that ensures that $t^\text{(ad)}_j \in \cT$ for all $j\in\cJ$. Additionally, based on the arrival deadline of DS $j\in\cJ$, we define the respective departure deadline at FC $i\in\cI$ as: 
\[
t^\text{(dd)}_{ij} = \max(t^\text{(ad)}_j - \delta_{ij}, 0),
\]
with the 0-th slot representing an invalid connection.  

The delay $\{\delta_{ij}\}$, product availability $\{a_{ik}\}$, and demand parameters $\{d_{jkt}\}$, as well as the arrival deadlines (cutoff) $\{t^\text{(ad)}_j\}$ and departure deadlines $\{t^\text{(dd)}_{ij}\}_{i\in\cI, j \in \cJ}$, are considered fixed and known. This implies that the behavior of the network is constant throughout the day and that there is a daily FC replenishment strategy, hence a product cannot go out of stock and the availability does not change in this time frame. One can envisage variations of this model by introducing stochastic perturbations and temporal components; we leave these extensions as future work.

\subsection{Decision Variables and Covered Demand}

\paragraph{Decision variables: } We introduce  the concept of \textit{last truck} of connection $(i,j)$, with $i\in \cl I, j\in \cl J$, as the latest timeslot prior or equal to $t^\text{(dd)}_{ij}$ that a truck can depart from FC $i$ towards DS $j$. Since there is no truck departure between the last truck and $t^\text{(dd)}_{ij}$, all orders made after the last truck cannot arrive before the cutoff time of the DS, where the last-mile vehicles will start the distribution. It is important to clarify that the demand is not covered solely by the last truck; for each connection $(i, j)$ there is a feasible schedule of trucks throughout the day that handles the requests arriving in earlier timeslots. The goal of our model is to optimize the schedule of the last truck for each active connection $(i,j)$ so as to maximize the DSs' demand that can meet the next-day delivery criteria. 

To that end, we define the binary variables $\vX\in \{0,1\}^{I\times J \times T}$, where $x_{ijt}$ decides if the last truck from FC $i\in\cI$ towards the DS $j\in\cJ$ departs at (the end of) timeslot $t\in\cT$. By definition, we can only have at most one last truck for each connection. Hence, the decision vector $\vx_{ij} \in \{0, 1\}^{T}$ for each \textit{active} connection $(i,j)$ will be of the type:
\[
\vx_{ij} = [0~~~0\hspace{-0.19in}\underbrace{1}_{\text{last truck departure}}\hspace{-0.19in}0~~~\dots ~~~0]^\top,
\]
which in this example indicates a departure of the last truck at the 3rd timeslot, while $\vx_{ij}=\vzero_T$ if the connection is \textit{inactive}, i.e., FC $i$ does not ship products to DS $j$. 

\paragraph{Covered demand:} The utility of a last-truck schedule $\vX$ is determined by the amount of demand it \emph{covers}. In particular, we say that demand $d_{jkt}$ can be served next-day (\emph{covered}) by FC $i\in \cl I$ if the following conditions are met: \emph{(i)} the FC stores this item, i.e., $a_{ik}=1$; and \emph{(ii)} the last truck departs at or after $t$, i.e., there is a timeslot $\tau \geq t$ where $x_{ij\tau}=1$. If a certain demand at some DS is covered by an FC, then clearly there is no need (or benefit) to cover it by additional FCs. In other words, the servicing benefit is not additive. In order to capture accurately this salient feature of the model, we define:
\begin{equation}
 1 - \prod_{i\in\cl I, \tau\geq t}\left(1 -a_{ik} x_{ij\tau} \right),
\end{equation}
which amounts to $1$ if demand $d_{jkt}$ is served by one or more FCs, and to $0$ if no FC that possess product $k$  has scheduled its last truck at or after $t$. Ultimately, we are interested in the aggregate demand that is covered by \emph{any} FC:
\begin{align}\label{eq:gx}
g(\vX)=\sum_{j\in\cl J, k\in\cl K, t\in\cl T}d_{jkt}
\left( 1 - \prod_{i\in\cl I, \tau\geq t} \left(1 - a_{ik} x_{ij\tau}\right)\right).
\end{align}

\begin{figure}[t]
\centering
  \includegraphics[width=0.6\linewidth]{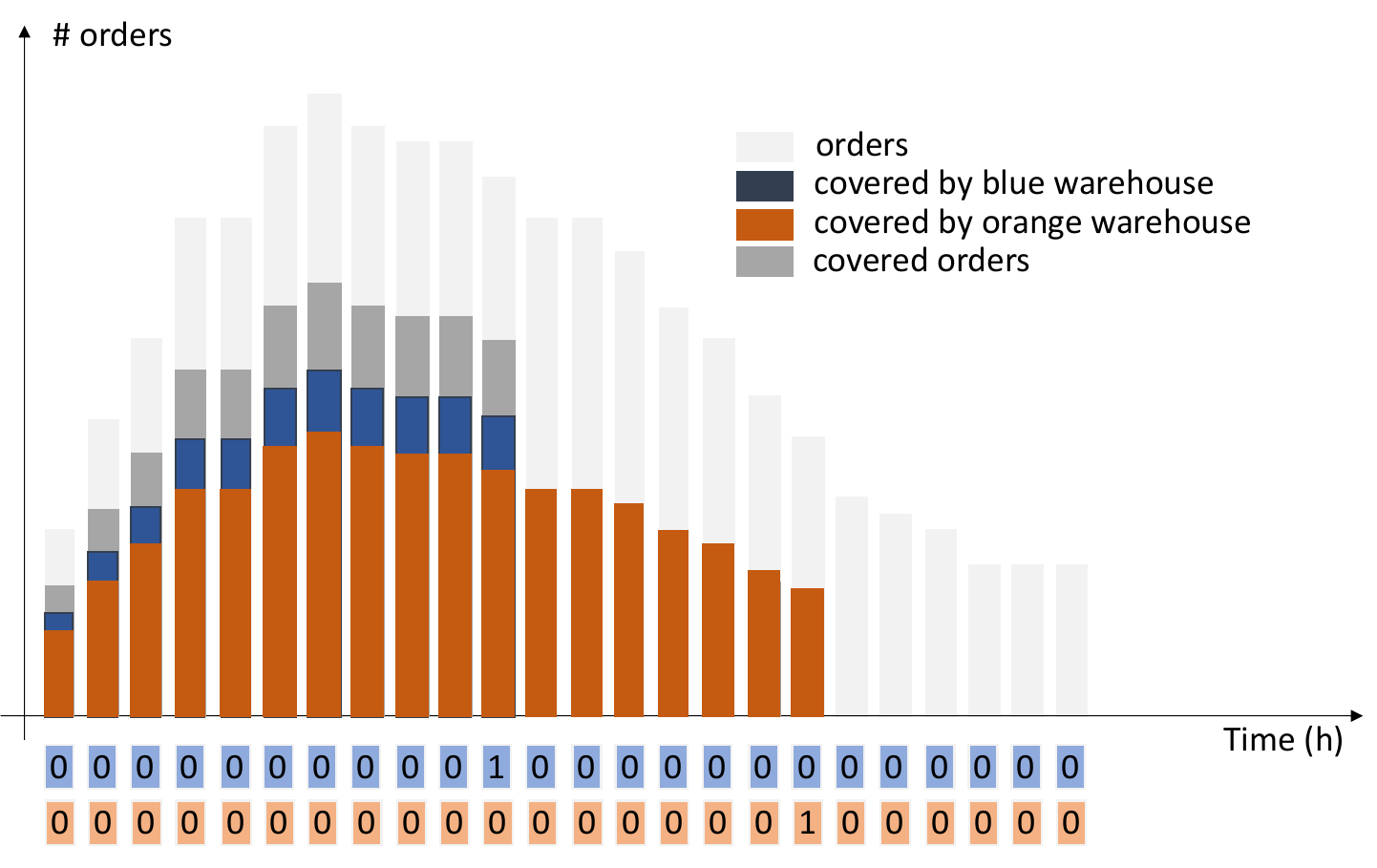}
  \caption{Coverage of orders with next-day delivery from individual and combined warehouses. Horizontal axis depicts the departure of the last-truck for each (of the two) FC, and vertical axis measures the number of orders that each FC can cover.}
  \label{fig:coverage}
\end{figure}

An example scenario is depicted in Figure \ref{fig:coverage} where two FCs can serve a DS with their last trucks departing at $t=11$ and $t=18$, as marked in the horizontal axis. The vertical axis shows the number of different orders that can be covered with next-day delivery for each possible warehouse combination, i.e., when using only the blue FC, when using only the orange FC, and when using both FCs (dark gray). The light-gray bars represent the total received orders for each time slot (here, granularity of hours is considered). Notice that the number of orders covered by both FCs are less than the sum of individually-covered orders, which is due to inventory overlap. Indeed $g(\vX)$ has the form of a coverage function, which is known to be monotone diminishing returns (DR) submodular  \citep{nemhauser1978analysis}.

Finally, note that we have restricted our analysis to networks that consist of FC-DS direct connections, while more complex networks with consolidation hubs \citep{wu-snd-ts23} are possible but left out of scope. Although the generalized mathematical model for these more complicated topologies is of its own interest, the additional benefit in NDD optimization is expected to be small. These intermediate nodes induce prohibitive loading/unloading and processing delays, as well as deviations from the shortest (direct) paths, hence they induce much earlier departures $t^\text{(dd)}_{ij}$ which in turn deteriorate the demand coverage.

\subsection{Optimization Model}\label{sec:problem}

Selecting $\vX$ to maximize the covered demand $g(\vX)$ is a challenging task due to the intricate structure of $g(\vX)$ and the operational requirements of the network. Firstly, we should not schedule tracks on connections that are inactive or on timeslots that yield delayed arrivals. To formalize this requirement, we define the respective set of timeslots where trucks \textit{cannot} be scheduled on $(i,j)$ as $\cl T_{ij}=\{t: t\in\cT, t > t^\text{(dd)}_{ij}\}$, and introduce the auxiliary binary vector $\mathbf{q}_{ij}\in \{0, 1\}^{T}$ with $q_{ijt}=1$ if $t\in \cl T_{ij}$ and $q_{ijt}=0$ otherwise. This yields the following constraint:
\begin{equation*}
    \vx_{ij}^\top \vq_{ij} = 0, \qquad \forall i\in\cI, \ \forall j\in\cJ.
\end{equation*}
Clearly, these sets can include other operating restrictions at the FCs and DSs. 

Due to its monotonicity, $g(\vX)$ would be maximized if we select $\vX$ such that all last trucks depart at their respective deadlines $t^\text{(dd)}_{ij}$. In practice this aggressive strategy is infeasible, as it would imply that all departing trucks leave the FC within a given time window, which in turn requires the FC to process a large number of packages within a small time interval. Such spiked production typically leads to mistakes, capacity outages, and delivery misses. To prevent this from happening, $\vX$ should satisfy some outbound (OB) capacity constraints that limit the number of trucks departing from each FC (towards different DSs) at any timeslot. Formally:
\begin{equation*}
\sum_{j\in\cl J} x_{ijt}\leq c^{\text{(ob)}}_{i}, \qquad \forall i \in \cl I, \ \forall t\in\cT,    
\end{equation*}
where $c^{\text{(ob)}}_{i}$ is a predetermined bound on the number of departing trucks per timeslot.

Similarly, if a large number of trucks arrive concurrently at the inbound stage of the DSs, it is likely to induce cost-inefficient processing delays and violate various space limitations. This, in turn, imposes an additional constraint on the truck schedule, which needs to ensure that no more than $c^{\text{(ib)}}_{j}$ trucks arrive at each DS $j\in \cl J$ at any given timeslot. Enforcing this constraint requires to account for the different travel delays of trucks on the various FC-DS paths. In particular, for each (arrival) timeslot $\tau \in \cl T$, we introduce the binary vector $\vp^{(\tau)}_{ij}\in \{0, 1\}^{T}$ with components $p^{(\tau)}_{ijt}=1$ for $t = \tau-\delta_{ij}$ and $0$ otherwise. That is, $\vp^{(\tau)}_{ij}$ indexes the slot when a truck needs to depart from $i\in\cI$ in order to arrive at $j\in\cJ$ at slot $\tau\in\cT$ (if such a valid slot exists). Hence, the DS inbound (IB) constraints can be expressed succinctly as:
\begin{equation*}
\sum_{i \in \cl I} \vx_{ij}^\top \vp^{(\tau)}_{ij}\leq c^{\text{(ib)}}_{j}, \qquad \forall j\in \cl J, \ \forall\tau\in \cl T.    
\end{equation*}
These outbound and inbound constraints force many trucks to depart earlier than their ideal slot $t^\text{(dd)}_{ij}$, and compound the maximum coverage last-truck schedule optimization. 

Putting the above together, our problem consists of selecting truck deadlines $\vX$ in order to maximize the covered next-day demand $g(\vX)$.

\textbf{\underline{NDD Optimization:}}
\begin{maxi!}|l|[2]
{\vX}
{g(\vX)={\sum_{j\in\cl J, k\in\cl K, t\in\cl T}
d_{jkt}\left( 1 - \prod_{i\in\cl I, \tau\geq t} \left(1 - a_{ik} x_{ij\tau}\right)\right)} \label{opt:obj_basic_mip}}
{\label{opt:basic_mip}}{}
\addConstraint{\mathbf{x}_{ij}^\top \mathbf{q}_{ij}}{=0,\, \quad}{\forall i \in \cl I, \ \forall j\in\cl J}{\label{opt:basic_mip_b}}
\addConstraint{\mathbf{x}_{ij}^\top \mathbf{1}_T }{\leq 1, \quad}{\forall i \in \cl I, \ \forall j\in\cl J}
{\label{opt:basic_mip_c}}
\addConstraint{\sum_{j\in\cl J} x_{ijt}}{\leq c^{\text{(ob)}}_{i}, \quad}{\forall i \in \cl I, \ \forall t\in\cT}{\label{opt:basic_mip_d}}
\addConstraint{\sum_{i\in\cl I} \mathbf{x}_{ij}^\top \vp^{(\tau)}_{ij} }{\leq c^{\text{(ib)}}_{j}, \quad}{\forall j\in\cl J, \ \forall\tau \in \cl T}{\label{opt:basic_mip_e}}
\addConstraint{\mathbf{x}_{ij}}{\in \{0, 1\}^T, \quad}{\forall i \in \cl I, \ \forall j\in\cl J}.{\label{opt:basic_mip_f}}
\end{maxi!}
Constraint \eqref{opt:basic_mip_b} ensures the operational requirements are satisfied as explained above, i.e., there are no delayed arrivals and each last middle-mile truck departs before $t^\text{(dd)}_{ij}$ so it can reach its destination $j\in\cJ$ before the injection cutoff $t^\text{(ad)}_j$. Constraint \eqref{opt:basic_mip_c} ensures each connection $(i,j)$ has at most one last truck; and \eqref{opt:basic_mip_d}, \eqref{opt:basic_mip_e} count --within a timeslot-- the number of last trucks that depart from an FC or arrive at a DS, respectively, and limit them to appropriate bounds. \eqref{opt:basic_mip_f} is the integrality constraint for variables $\{x_{ijt}\}_{i\in\cI, j\in \cJ, t\in \cT}$.

\subsection{Problem Variations}

In this section, we present two problem instances that arise when we relax some of the operational constraints of problem \eqref{opt:basic_mip}, namely the inbound (IB) and outbound (OB) restrictions. For notation convenience, let us define the set of basic constraints that are common in all problem instances, as follows:
\begin{align*}
    &    \cX \doteq \Big\{\vX : \vX \in [0,1]^{I\times J \times T}, \eqref{opt:basic_mip_b}, \eqref{opt:basic_mip_c} \Big\},
\end{align*}
and the sets that include additionally outbound and inbound constraints for both the continuous relaxation and integer problem variants, as:    
\begin{align*}
&    \cX^{\text{(ob)}} \doteq \Big\{\vX : \vX \in \cX, \eqref{opt:basic_mip_d}  \Big\},\quad    &&\cX^{\text{(ib)}} \doteq \Big\{\vX : \vX \in \cX, \eqref{opt:basic_mip_e} \Big\},\\
&    \cX^{\text{(ob)}}_{\text{int}} \doteq \Big\{\vX : \vX \in \cX, \eqref{opt:basic_mip_d},\eqref{opt:basic_mip_f} \Big\},\quad
  &&\cX^{\text{(ib)}}_{\text{int}} \doteq \Big\{\vX : \vX \in \cX, \eqref{opt:basic_mip_e}, \eqref{opt:basic_mip_f}  \Big\}.
\end{align*}

Now, we can write \eqref{opt:basic_mip} in the following compact form:
\begin{maxi}|l|[2]
{\vX \in \cX^{\text{(ob)}}_{\text{int}}\cap\cX^{\text{(ib)}}_{\text{int}}}
{g(\vX),}
{\label{opt:P}}
{\cP: \quad }
\end{maxi}
and the two problem instances with partial constraints as 
\begin{maxi}|l|[2]
{\vX \in \cX^{\text{(ob)}}_{\text{int}}}
{g(\vX),}
{\label{opt:P_ob}}
{\cP^{\text{(ob)}}: \quad }
\end{maxi}

\begin{maxi}|l|[2]
{\vX \in \cX^{\text{(ib)}}_{\text{int}}}
{g(\vX).}
{\label{opt:P_ib}}
{\cP^{\text{(ib)}}: \quad }
\end{maxi}

The study of these problems not only facilitates the analysis of $\cP$ as it will become clear in later sections, but is also of practical interest as some or e-retailers might not have all the operational inbound and outbound restrictions of $\cP$. In the following, we characterize the properties of each problem variation (Section \ref{sec:theoretical}) and design a dedicated solution strategy for each instance based on its special properties (Sections~\ref{sec:ob}-\ref{sec:iob}).

%% file: complexity-v3.tex
\section{Theoretical analysis of $\cP$ and its Variants}
\label{sec:theoretical}

In this section we analyze the theoretical properties of the different problem variants. The goal of this analysis is two-fold. First, to characterize the complexity (or hardness) of the problems at hand and secondly to identify directions for solution methods with optimality guarantees or promising practical performance. 

\subsection{Complexity Analysis}
\label{sec:complexity}

We start with an analysis of the worst-case complexity of the NDD maximization problems. Specifically, we are able to polynomially transform some simple instances (with small number of time slots) of $\cP^{\text{(ob)}}$ into the femtocaching problem \citep{femtocaching}, and similarly transform some simple instances of $\cP^{\text{(ib)}}$ into the maximum coverage problem \citep{feige}. These transformations lead us to the following theorem.

\begin{theorem}
\label{the:complexity}
The worst-case complexity of NDD problems is characterized as follows:
\begin{itemize}
\item $\cP^{\text{(ob)}}$ is NP-hard.
\item $\cP^{\text{(ib)}}$ is NP-hard to approximate to $\geq\left(1-1/e\right)\cdot \texttt{opt}$.
\item $\cP$ is NP-hard to approximate to $\geq\left(1-1/e\right)\cdot \texttt{opt}$.
\end{itemize}
\end{theorem}

\textbf{Proof.} The NP-hardness of $\cP^{\text{(ob)}}$ is based on a polynomial reduction from the femtocaching problem, see Appendix \ref{appendix:reduction_ob}. The inapproximability result of $\cP^{\text{(ib)}}$ follows from a polynomial reduction from the maximum coverage problem with cardinality constraint, see Appendix \ref{appendix:reduction_ib}. The result of $\cP$ follows directly as it generalizes $\cP^{\text{(ib)}}$.

\subsection{Problem Properties}
\subsubsection{Properties of Objective Function}   

We start with the objective function which is common in all problems. Following the submodular optimization terminology, we define the abstract set of elements, $\cl V=\left\{ v_{ijt}\right\}_{i\in \cl I,\ j\in \cl J, \, t\in \cl T}$, where $v_{ijt}$ represents the decision to schedule a truck from FC $i\in\cI$ to DS $j\in\cJ$ at slot $t\in\cT$. To put things in context, each element $v_{ijt}$ can be interpreted as a subset of items from a universe $\cl D=\{d_{jkt}\}_{j\in\cl J, k\in \cl K, t\in\cl T}$, which includes the demand values $d_{jkt}$ from all DSs, products and slots. Further, we denote with $\cl S \subset \mathcal V$ the set that consists of the elements induced by the assignment matrix $\vX$; we abuse the notation and use $g(\cl S)$ and $g(\vX)$ interchangeably. Based on this interpretation, the objective function can be equivalently defined as a \emph{set function} and the next lemma defines its properties. 

\begin{lemma} \label{lem:submodular}
The set function $g:2^{\cl V}\rightarrow \mathbb R_+$ is monotone submodular increasing.
\end{lemma}
\textbf{Proof.} See Appendix  \ref{appendix:proof-submodular-function}.

We also note that the alternative formulation,
\[
g(\vX)=\sum_{j\in\cl J, k\in\cl K, t\in\cl T}
d_{jkt}\left( 1 - \prod_{i\in\cl I, \tau\geq t} \left(1 - a_{ik} x_{ij\tau}\right)\right)\]
retains this property, and remark that $g(\cdot)$ is a coverage function, e.g., see \cite{femtocaching} for a recent discussion.

\subsubsection{Properties of Constraints}   

We next analyze the structure of constraints. We first observe that we can drop constraint \eqref{opt:basic_mip_c} since the last truck covers the demand that all earlier trucks can cover. In terms of coverage, this means that for any two elements $v_{ijt}$ and $v_{ijt'}$ with $t<t'$, $v_{ijt'}$ covers a superset from the universe $\cl D$ compared to $v_{ijt}$, hence the latter is non-contributing if $v_{ijt'}$ is included in $\vX$. We define $\bar{\cX}^{\text{(ob)}}_{\text{int}} = \cX^{\text{(ob)}}_{\text{int}} / \eqref{opt:basic_mip_c}$ and $\bar{\cX}^{\text{(ib)}}_{\text{int}} = \cX^{\text{(ib)}}_{\text{int}} / \eqref{opt:basic_mip_c}$, i.e., the constraint sets without \eqref{opt:basic_mip_c}. With that in mind, we arrive at the next result.
\begin{lemma} \label{lem:partition-intersection-all}
The set of constraints $\bar{\cX}_{\text{int}} = \bar{\cX}^{\text{(ob)}}_{\text{int}}\cap\bar{\cX}^{\text{(ib)}}_{\text{int}}$ can be expressed as the intersection of two partition matroids, and this intersection does not constitute a matroid.
\end{lemma}
\textbf{Proof}. See Appendix  \ref{appendix:proof-matroid}.

The main idea is that the ground set $\cl V$ can be partitioned in two ways, and any solution should be \emph{independent} with respect to both partitions. First, $\cl V$ can be expressed as the union of $T\cdot I$ disjoint subsets (i.e., {partitions}) $\cl V_{it}$, $i\in \cl I, t\in \cl T$, each representing the last trucks which can depart from FC $i\in\cI$ towards all DSs at slot $t\in\cT$, and which cannot exceed the FC's outbound capacity. At the same time, $\cl V$ can be expressed as the union of $J\cdot T$ disjoint subsets, $\cl V_{jt}$, $j\in \cl J, t\in \cl T$, representing the possible trucks arriving at each DS $j\in\cJ$ from any FS at slot $t\in\cT$, and which cannot exceed the DS's inbound capacity. These two families of sets constitute partition matroids. The next corollary follows directly from the above Lemma.
\begin{corollary}
Constraint sets $\bar{\cX}^{(ob)}_{\text{int}}$ and $\bar{\cX}^{(ib)}_{\text{int}}$ are partition matroids. 
\end{corollary}
Indeed, each of them can be defined with the help of one of the above partition matroids.

\subsection{Guarantees}
\label{ssec:guarantees}

One can observe that $\cP^{(ib)}$ can be decomposed into $J$ subproblems, one per DS $j\in \mathcal J$, since its objectives and constraints can be readily grouped per DS. This separation does not alter the properties of the problem but it does reduce its dimension and allows for their parallel solution. This problem has been studied extensively in the context of submodular optimization. \cite{chekuri-max-budget} showed that a greedy algorithm can achieve a ratio of $1/2$, and algorithms with the optimal (tight) ratio of $1-1/e$ were presented in \citep{ageev2004pipage, calinescu2011maximizing} using pipage rounding or local search \citep{filmus-local-2012}. In this work we build our solution strategy using the greedy and also the pipage rounding algorithms, where we also exploit the inherent decomposability of the problem.

Similar strategies, albeit without the decomposability, can be followed for $\cP^{(ob)}$, which maximizes the same function over a different partition matroid. Hence, the greedy algorithm and pipage rounding guarantee approximations of $1/2$ and $1-1/e$, respectively. Finally, regarding the overall problem with inbound and outbound constraints, we have the following direct result. 
\begin{corollary}
Problem $\cP$ is a submodular maximization problem subject to intersection of two partition matroids. 
\end{corollary}
Submodular maximization over multiple matroids is discussed in \citep{lee-mathOR2010}. In \citep{nemhauser1978analysis} it was shown that the classical greedy algorithm achieves a $1/(k+1)$ approximation for the problem of maximizing a monotone submodular function subject to $k$ matroid constraints. In the special case of partition matroids, as in $\cP$, \cite{lee-siam-2010} devised a local search algorithm that achieves an improved approximation of $1/(k+\epsilon)$ for any $k\geq 2$ and $\epsilon > 0$. This result was subsequently extended to general matroids by \cite{lee-mathOR2010}. This means that, for the problem at hand, we can use a similar local search algorithm to achieve $1/2+\epsilon$, or even the simplest greedy algorithm to achieve $1/3$. However, the  solution we follow is more involved. Namely, we choose a partial Lagrange relaxation, cf. \citep{fischer-lagrange}, and a dual-ascent with the optimal step proposed in \citep{polyak-step}, where in each iteration we solve a modified variant of $\cP^{(ob)}$ or $\cP^{(ib)}$. Using a wealth of experiments with realistic datasets and scenarios, we find this approach to yield near-optimal performance with minimal computation overheads -- thus exceeding substantially the above guarantees.

%% file: OB.tex
\section{Solving the Partial Problems $\cP^{\text{(ob)}}$ and $\cP^{\text{(ib)}}$}
\label{sec:ob}

We start with problem $\cP^{\text{(ob)}}$ (defined in \eqref{opt:P_ob}), i.e., the variant of $\cP$ where only the inbound constraints are dropped. 
In the following, we describe different approaches to tackle $\cP^{\text{(ob)}}$. As we will show, there is an alternative way to define $\cP^{\text{(ob)}}$ as an ILP that, in conjunction with its structure, allows to use pipage rounding \citep{ageev2004pipage} that returns an integer solution with optimality guarantees.  It is important to highlight the applicability of pipage rounding in $\cP^{\text{(ob)}}$ is due to the properties of the objective and constraints. We further describe a greedy algorithm, which is a common approach in submodular maximization problems. In Section \ref{ssec:ob_eval} we will explore the pros and cons of these approaches, and compare them with an exact solution obtained using state-of-the-art commercial solvers \citep{fico-xpress, gurobi}.

\subsection{Pipage Rounding} \label{sec:pipage}

\subsubsection{Overview}

Pipage is an iterative rounding algorithm that starts with an initial fractional solution $\vX$  and iteratively  converts it to integral without decreasing the objective value. Although the algorithm itself is straightforward, its application is problem-specific. The first requirement of pipage is an initial well-performing fractional solution $\vX$. The objective $g(\cdot)$ of $\cP^{\text{(ob)}}$ (and $\cP$ in general) has the property that it obtains its maximum value in integral points, even in the absence of integrality constraints \cite[Ch. 44]{schrijver2003combinatorial}. Consequently, even if we relax the integrality constraints of $\cP^{\text{(ob)}}$ the remaining problem is a non-linear maximization which is still NP-hard. To overcome this, the main idea (used extensively in pipage algorithms) is to reformulate the initial problem $\cP^{\text{(ob)}}$ with a new objective $f(\cdot)$ that is point-wise related to $g(\cdot)$ and is easy to optimize its continuous relaxation \cite{ageev2004pipage}. The solution of the  continuous relaxation of the reformulated  problem provides the initial fractional solution $\vX^{(0)}$ for the pipage rounding scheme. 

Given an initial solution $\vX^{(0)}$, we can apply the iterative pipage rounding algorithm, where at each step we obtain $\vX^{(\ell+1)}$ from $\vX^{(\ell)}$, for $\ell=0,1,\dots,L-1$, until convergence to $\vX^{(L)}$, where $L$ is the number of iterations. By construction of the pipage algorithm, $\vX^{(\ell+1)}$ has (at least) one less fractional variable compared to $\vX^{(\ell)}$, that is converted into integer, while it ensures that the newly obtained solution $\vX^{(\ell+1)}\in \cX^{\text{(ob)}}$ is feasible (only violating the integrality constraints), and does not deteriorate  the value of the objective, i.e., $g(\vX^{(\ell+1)})\geq g(\vX^{(\ell)})$. 

Some key ingredients that are problem specific and we need to study in such a method are: \emph{(i)} what is the new objective $f(\cdot)$ and how is it related to $g(\cdot)$; and \emph{(ii)} how to compute a pipage step that keeps the solution feasible, does not decrease the objective and converts at least one more variable to integer. 
In Section \ref{sec:reformulation} we provide details of the problem reformulation and its relaxation, while we prove a point-wise relation between $g(\cdot)$ and $f(\cdot)$. In Section \ref{sec:approximation} we introduce how a general pipage step works and present the complete approximation algorithm. For the interested reader we defer the discussion on how to find such a step for our problem in Appendix \ref{sec:pipage_step_ob}, as it follows standard arguments \citep{ageev2004pipage}. Finally, in Section \ref{sec:pip_strategies} we discuss how the order of pipage steps matters and provide possible strategies.

\subsubsection{Problem Reformulation.}   
\label{sec:reformulation}

We first introduce a modified coverage function:
\begin{equation}
\label{eq:fx}
f(\vX)=\sum_{j\in\cl J, k\in\cl K, t\in\cl T} d_{jkt} \min\left(1, \sum_{i\in\cI, \tau \geq t} a_{ik} x_{ij\tau}\right).
\end{equation}
The term $\min\left(1, \sum_{i\in\cI, \tau \geq t} a_{ik} x_{ij\tau}\right)$ becomes $1$ if demand $d_{jkt}$ is covered by at least one FC, and $0$ otherwise. It turns out that $g(\cdot)$ and $f(\cdot)$ are not identical, but the following lemma shows that they are point-wise related in the entire domain.
\begin{lemma}
\label{lemma:rho}
For all $\vX \in \cX^{\text{(ob)}}
$, we have $g(\vX) \geq \rho(mT) \cdot f(\vX)$ where $\rho(x) \!=\! 1\! - (1\! - 1/x)^x$ and $m$ is the maximum number of FC connections that a DS can have.
\end{lemma}
\textbf{Proof.} See Appendix \ref{app:proof_lemma_rho}.

Furthermore,  $g(\cdot)$ and $f(\cdot)$  are equal on integers, hence the optimization problem 
\begin{maxi}|l|[2]
{\vX \in \cX^{\text{(ob)}}_{\text{int}}}
{f(\vX)}
{\label{opt:fx_mip}}
{\cP_f^{\text{(ob)}}: }
\end{maxi}
has the same optimal solution, $\vX^\star$, and optimal value, $\texttt{opt}$, as $\cP^{\text{(ob)}}$. Obviously, $\cP_f^{\text{(ob)}}$ is also NP-hard and has a nonlinear objective. This reformulation has two advantages. First, it is straightforward to rewrite $\cP_f^{\text{(ob)}}$ as an integer linear program (ILP):
\begin{maxi}|l|[2]
{\vX, \vY}
{\sum_{j\in\cl J, k\in\cl K, t\in\cl T} d_{jkt} y_{jkt}}
{\label{opt:fx_ilp}}
{\cP_{f_{ILP}}^{\text{(ob)}}: }
\addConstraint{y_{jkt}}{\leq \sum_{i\in\cI, \tau \geq t} a_{ik} x_{ij\tau}, \quad}{\forall j\in\cJ,\ \forall k\in\cK,\ \forall t\in\cT}
\addConstraint{y_{jkt}}{\leq 1, \quad}{\forall j\in\cJ,\ \forall k\in\cK,\ \forall t\in\cT}
\addConstraint{\vX}{\in\cX_{\text{int}}^{\text{(ob)}}}{},
\end{maxi} 
where we have introduced the auxiliary (continuous) variables $\vY\in [0, 1]^{|J\times K \times T}$ and replaced the nonlinear $\min(\cdot)$ operator with constraints. Second, although $\cP^{\text{(ob)}}$ and $\cP_{f_{ILP}}^{\text{(ob)}}$ are equivalent, in the ILP we can discard the integrality constraints of $\cP_{f_{ILP}}^{\text{(ob)}}$ and solve its continuous relaxation, i.e., the following LP:
\begin{maxi}|l|[2]
{\vX, \vY}
{\sum_{j\in\cl J, k\in\cl K, t\in\cl T} d_{jkt} y_{jkt}}
{\label{opt:fx_lp}}
{\cP_{f_{LP}}^{\text{(ob)}}: }
\addConstraint{y_{jkt}}{\leq \sum_{i\in\cI, \tau \geq t} a_{ik} x_{ij\tau}, \quad}{\forall j\in\cJ,\ \forall k\in\cK,\ \forall t\in\cT}
\addConstraint{y_{jkt}}{\leq 1, \quad}{\forall j\in\cJ,\ \forall k\in\cK,\ \forall t\in\cT}
\addConstraint{\vX}{\in\cX^{\text{(ob)}}}{}.
\end{maxi}

We denote with $f(\vX^{(0)})$ the optimal value of $\cP_{f_{LP}}^{\text{(ob)}}$, where $\vX^{(0)}$ is the optimal (fractional) solution. 
Since $f(\vX^{(0)})$ is the optimal value of the relaxed version of $\cP_{f_{ILP}}^{\text{(ob)}}$ (and equivalently $\cP_{f}^{\text{(ob)}}$), which has the same optimal (integral) value as $\cP^{\text{(ob)}}$, $\texttt{opt}$, it holds:
\begin{equation}
\label{eq:opt_f}
\texttt{opt} = g(\vX^\star) = f(\vX^\star) \leq f(\vX^{(0)}),
\end{equation}
i.e., $f(\vX^{(0)})$ is an upperbound on the optimal value.

\subsubsection{Approximation Algorithm.}   
\label{sec:approximation}

We next introduce the concept of \emph{pipage step} that is used in the approximation algorithm.

\begin{definition}[Pipage step]\label{def:pipage}
Given any non-integral solution $\vX^{(\ell)} \!\in\! \cX^{\text{(ob)}}$, a pipage step provides another (possibly non-integral) solution $\vX^{(\ell + 1)}$ with properties:
\begin{enumerate}[(i)]
    \item $\vX^{(\ell + 1)} \in \cX^{\text{(ob)}}$, hence it satisfies  all but the integrality constraints.
    \item $\vX^{(\ell + 1)}$ is not decreasing the objective, i.e., $g(\vX^{(\ell + 1)}) \geq g(\vX^{(\ell)})$.
    \item $\vX^{(\ell + 1)}$ has at least one more integer value than $\vX^{(\ell)}$.
\end{enumerate}
\end{definition}

\begin{algorithm}[t]
{\begin{small}
\caption{Pipage rounding for $\cP^{\text{(ob)}}$}\label{alg:main_algo}
\begin{algorithmic}
\State $\vX^{(-1)} \gets \boldsymbol{0}$
\State $\vX^{(0)} \gets$ the solution of $\cP_{f_{LP}}^{\text{(ob)}}$ 
\State $\ell \gets 0$ 
\While{$\vX^{(\ell)} \neq \vX^{(\ell - 1)}$}
	\State $\vX^{(\ell + 1)} \gets \texttt{Pipage-step}(\vX^{(\ell)})$
	\State $\ell\gets \ell+1$
\EndWhile
\end{algorithmic}
\end{small}}
\end{algorithm}

Now, we can introduce Algorithm~\ref{alg:main_algo} which iterates pipage steps until convergence. The algorithm terminates when the $\texttt{Pipage-step}$ subroutine returns the same solution, i.e., if solution $\vX^{(L)}$ has only integer elements. Hence, the algorithm stops in $L$ steps, with $L\leq I\cdot J \cdot T$. We discuss how to find a pipage step in Appendix \ref{sec:pipage_step_ob}, while $\texttt{Pipage-step}$ is given in Algorithm \ref{alg:step}.

Putting the above together, we can derive the following result.
\begin{lemma}
\label{lemma:approx}
For every instance of $\cP^{\text{(ob)}}$, Algorithm \ref{alg:main_algo} stops after finite steps and returns an integer solution $\vX^{(L)} \in \cX^{\text{(ob)}}_{\text{int}}$ that ensures 
\[
g(\vX^{(L)}) \geq \rho(mT)\cdot\texttt{opt},
\]
where $\rho(x)= 1 - (1 - 1/x)^x$, with $\rho(\infty)=\lim_{x\to\infty}\rho(x)=1-1/e\approx 0.632$, and $m$ is the maximum number of inbound connections to any destination node in $\cJ$. 
\end{lemma}
\textbf{Proof.}  See Appendix \ref{app:proof_lemma_approx}.

\subsubsection{Pipage Step Ordering Strategies}   \label{sec:pip_strategies}

The pipage step in Algorithm \ref{alg:main_algo} selects randomly two non-integral variables to update during each iteration. This is a standard approach in pipage algorithms, yet the selection strategy (or, rule) has important implications that are typically overlooked. Here, we provide a set of such rules and study experimentally their impact on the final objective value. To exemplify, consider a toy scenario with two FCs that cover partially the demand of a certain DS. Updating first FC $1$ can cover fully the demand of the DS and this, in turn, will render FC $2$ coverage obsolete, which will be thus removed in a subsequent pipage step. The opposite can also happen if we update first FC $2$. With a similar reasoning, it is easy to see that even the order of variable updates within a given FC (i.e., towards its various DS connections) might affect the result. Unfortunately, finding the rule that achieves the highest possible objective value requires navigating a vast search space; hence we will resort to local search methods. We present three such rules with different speed-performance trade-offs, that determine the FC-selection, i.e., update order of FCs, and select randomly the variables within each FC. We stress at this point that, independently of the selection rule, the algorithm terminates after $L$ steps with a solution that is guaranteed to satisfy $\geq (1-1/e)\cdot \texttt{opt}$.

\paragraph{Order once and fix (OOF):} In this strategy, we initially apply pipage rounding in parallel to all FCs. This way we apply $I$ parallel updates, where in the $i$-th update the variables of FC $i$ are integral, i.e., $\vX_i\in\{0, 1\}^{J\times T}$, and the variables of FCs $\cI\setminus\{i\}$ are equal to the fractional solution $\vX^{(0)}$. With a slight abuse of notation, denote by $\vX^{(0, i)}$ the complete (fractional) solution with only FC $i$ updated. The incremental gain of the pipage update can be computed as $g(\vX^{(0, i)}) - g(\vX^{(0)}) \geq 0$. The order of the FCs stems naturally based on the value of their gain. 

Now, we update sequentially the initial fractional solution $\vX^{(0)}$ at an FC level and in the predefined order with the already computed pipage updates. At each step we recompute the gain (since it has changed due to the previous steps) but keep the originally computed pipage updates even if the gain drops (unless becomes negative). This strategy has the major advantage of being fast, since all FCs are updated only once and in parallel. On the other hand, OOF has reduced performance because the parallel updates do not take into account information from the already-updated FCs.

\paragraph{Order once and update (OOU):} This strategy is similar to OOF but addresses the performance issue. In OOU we use the fixed ordering of FCs obtained in the same way as before, but at each step we recompute the pipage rounding at the corresponding FC, hence we take into account all previous steps. This strategy is slower than OOF since it is equivalent to a sequential rather than a parallel update, but it improves the objective.

\paragraph{Order at each step (OES):} In this strategy we order the FCs at each step based on their updated incremental gains in previous steps. Thus, although there is an initial order, it is not fixed and it is perpetually updated. This method is similar to a local search at each step, is slower than OOF and OOU, but yields the highest objective values.

\subsection{Greedy Algorithm}
\label{ssec:greedy_ob}

Greedy algorithms are one of the main tools for tackling submodular maximization problems \citep{krause2014submodular}. They provide solutions with optimality guarantees, which are often exceeded in practice, and they have low overheads and small runtimes. We describe below such a greedy algorithm specifically for problem $\cP^{\text{(ob)}}$. We remind the reader that, since this is a sumbodular maximization problem subject to one partition matroid, we know from the discussion in Section \ref{ssec:guarantees} that this algorithm will achieve a solution that is at least $1/2$ of the optimal. 

The implementation of the greedy algorithm is simple: for all possible FC-DS connections we calculate the incremental gain we get if we schedule a truck as late as possible, namely at $t^\text{(dd)}_{ij}$, if there are no other operational constraints. Then, we initialize a priority queue with all these connection-timeslot pairs based on their gain. At each step, we pick the top pair from the priority queue. If the resulting schedule is still feasible (based on the previously added pairs) and its gain is still higher than the second element of the queue, we assign a truck at that timeslot. Otherwise, we find the next feasible slot of the connection and/or update the gain and place the pair back to queue. The algorithm terminates when no more trucks can be assigned.

%% file: IB.tex
\subsection{Pipage Solution and Decoupling of Problem $\cP^{\text{(ib)}}$}\label{sec:ib}

We shift our focus to problem $\cP^{\text{(ib)}}$, defined in \eqref{opt:P_ib}, i.e., the variant of $\cP$ where only the inbound operational constraints are imposed and the outbound constraints are dropped.  A straightforward approach to solve $\cP^{\text{(ib)}}$ is to follow the same strategy as in Section~\ref{sec:pipage}, and use pipage rounding to construct an approximation algorithm. The resulting procedure is similar to Algorithm \ref{alg:main_algo} and therefore omitted. The details about the pipage step subroutine for problem $\cP^{\text{(ib)}}$ are provided in Appendix \ref{sec:pipage_step_ib}. 

Due to its structure, $\cP^{\text{(ib)}}$ decouples per $j\in\cJ$. For a fixed $j\in\cJ$, one needs to solve: 
\begin{maxi}|l|[2]
{\vX_j}
{g(\vX_{j})={\sum_{k\in\cl K}\sum_{t\in\cl T}
d_{jkt}\left( 1 - \prod_{i\in\cl I, \tau \geq t}\left(1 - a_{ik} {x_{ij\tau}}\right)\right)}}
{\label{opt:basic_j_mip}}
{\cP_j^{\text{(ib)}}: }
\addConstraint{{\mathbf{x}_{ij}}^\top \mathbf{q}_{ij}}{=0,\, \quad}{\forall i \in \cl I}{}
\addConstraint{{\mathbf{x}_{ij}}^\top \mathbf{1}_T }{\leq 1, \quad}{\forall i \in \cl I}
{}
\addConstraint{\sum_{i\in\cl I} {\mathbf{x}_{ij}}^\top \vp^{(\tau)}_{ij} }{\leq c^{\text{(ib)}}_{j}, \quad}{\forall\tau \in \cl T}{}
\addConstraint{\mathbf{x}_{ij}}{\in \{0, 1\}^T, \quad}{\forall i \in \cl I}{}.
\end{maxi}
This is particularly important for pipage because it implies the updates of one DS do not affect the downstream decisions on other DSs. Thus, since the DSs are independent, we can apply pipage updates to all DSs in parallel, i.e., strategy OOF, as ordering does not play any role.

Furthermore, the problem decoupling is advantageous not only for pipage but for the direct solver solution as well. Instead of a high-dimensional problem, a solver needs to optimize \textit{in parallel} $J$ medium-sized problems (in the order of $10^3 - 10^4$ binary variables for realistic networks), which decreases significantly the overall running time. Note that the solver optimizes the corresponding ILP version of \eqref{opt:basic_j_mip}, derived in a similar manner as in Section \ref{sec:reformulation}, by replacing $g(\cdot)$ with $f(\cdot)$. Finally, given the properties of the problem (see Section \ref{ssec:guarantees}), we can also employ a greedy algorithm for solving $\cP^{\text{(ib)}}$.

%% file: IOB.tex
\section{Solution Methods for $\cP$}\label{sec:iob}

In the previous sections we described the properties and solution approaches for $\cP^{\text{(ob)}}$ and $\cP^{\text{(ib)}}$. We found that for $\cP^{\text{(ob)}}$ and $\cP^{\text{(ib)}}$ there is an efficient approximation algorithm (pipage rounding) which solves the problem with optimality guarantees and, in practice, performs even better than those guarantees. Additionally, $\cP^{\text{(ib)}}$ can be decomposed per DS, hence its solution is equivalent to solving $J$ smaller problems in parallel. 

In this section we study the complete problem $\cP$ where both outbound and inbound operational constraints are imposed. This problem is particularly challenging since it inherits the drawbacks of both $\cP^{\text{(ob)}}$ and $\cP^{\text{(ib)}}$, i.e., it does not decouple due to the outbound constraints, and pipage cannot be applied due to the presence of both constraints. Our strategy for tackling $\cP$ is to perform a partial Lagrange relaxation on the set of OB or IB constraints, and thus obtain a modified problem that has similar structure with the ones studied in the previous sections. This approach leads eventually to an iterative algorithm, a dual descent method in particular, where at each step we solve the Lagrange problem (either with pipage or with a solver, due to decoupling), and update accordingly the dual variables using the subgradients obtained from the Lagrange maximization. This type of Lagrangian heuristics have been proven very successful for a range of NP-hard operation research problems, see \citep{fischer-lagrange}. We find that this is the case indeed also for the problem at hand.

\subsection{Lagrangian Relaxation of IB Constraints}
\label{ssec:lag_ib}

We start with an iterative algorithm for $\cP$ by relaxing the IB constraints, where we denote with $\cP^{L{(\text{ib})}}$ the following problem:
\begin{maxi}|l|[2]
{\vX, \vlambda}
{{g(\vX) + \sum_{j\in\cl J}\sum_{\tau\in\cl T} \lambda_{j\tau}\left(c^{\text{(ib)}}_{j} - \sum_{i\in\cl I} \mathbf{x}_{ij}^\top \vp^{(\tau)}_{ij}\right)} }
{\label{opt:lag_ib}}{\cP^{L{(\text{ib})}}: }
\addConstraint{\vX\in\cX^{\text{(ob)}}_{\text{int}},}{}
\end{maxi}
where $\vlambda \in \bR_+^{I\times T}$ are the dual variables of the relaxed constraint. Note that this problem is identical to $\cP^{\text{(ob)}}$ (defined in \eqref{opt:P_ob}) with the additional Lagrangian terms in the objective. 

Section \ref{sec:ob} detailed how pipage applies to $\cP^{\text{(ob)}}$, which was enabled by the properties of its objective and constraints. For fixed $\vlambda$, the additional linear terms in the objective do not break the applicability of pipage. Therefore, the process is identical to the one presented in Algorithms \ref{alg:main_algo} and \ref{alg:step}. We omit the proof as it is a trivial extension of the analysis for $\cP^{\text{(ob)}}$. In a nutshell, the difference is only in the objective and hence we only need to examine the convexity of the pipage directions. Indeed, a convex direction for $\cP^{\text{(ob)}}$ is also convex for $\cP^{L{\text{(ib)}}}$ as the additional linear terms preserve convexity.

The pipage applicability on the above formulation gives rise to an efficient dual descent algorithm. Define as $\cP_{f_{LP}}^{L{(\text{ib})}}$ the LP version of \eqref{opt:lag_ib}, i.e., reformulate $g(\cdot)$ with $f(\cdot)$ and relax the integrality constraints (see also Section \ref{sec:reformulation}). At each iteration, with fixed dual variables $\vlambda$, we solve $\cP_{f_{LP}}^{L{(\text{ib})}}(\vlambda)$ to obtain an initial (fractional) solution and then apply pipage to convert it to integral. Accordingly, we update $\vlambda$ based on the new integral solution $\vX$, using the Polyak step. We repeat until a convergence criterion is satisfied, or a maximum number of iterations is reached. 

% As we saw in Section \ref{sec:ob}, in order to apply pipage we require an initial solution. We can get this in exactly the same way as for $\cP^{\text{(ob)}}$, i.e., by reformulating the objective using $f(\cdot)$ (which gives an ILP) and then relaxing the integrality constraints. The corresponding LP (for fixed $\vlambda$) is
% \begin{maxi}|l|[2]
% {\vX}
% {{f(\vX) + \sum_{j\in\cl J}\sum_{\tau\in\cl T} \lambda_{j\tau}\left(c^{\text{(ib)}}_{j} - \sum_{i\in\cl I} \mathbf{x}_{ij}^\top \vp^{(\tau)}_{ij}\right)}}
% {\label{opt:f_LP_lag_ib}}{\cP_{f_{LP}}^{L{(\text{ib})}}(\vlambda): }
% \addConstraint{\vX\in\cX^{\text{(ob)}}.}{}
% \end{maxi}

Since the IB constraints are relaxed, there is no guarantee that the final solution will be feasible \citep{fischer-lagrange}. To address this issue, we propose a fast and efficient approach that recovers feasibility of the IB constraints at each iteration. The idea is simple: we find the (usually small) subset of trucks that, if removed (i.e., assign the corresponding entries of $\vX$ to 0) will alleviate all the constraint violations. Then, we run a greedy algorithm (as the one in Section \ref{ssec:greedy_ob}) only on the connections that trucks were removed (with all the feasible trucks fixed) in order to place back these trucks in a locally-optimal fashion, while retaining feasibility. Alternatively, one can run this greedy feasibility correction only at the last iteration. However, in practice we want to keep the best solution over all iterations, i.e., the solution with the highest objective that is also feasible, and therefore we opt for the first option. We stress that the computational overhead of the greedy feasibility routine is minimal compared to pipage. All the steps are summarized in Algorithm \ref{alg:iterative_pip_ob}, where parameter $N$ defines the number of maximum iterations and $a_{\ell}$ is the step size for the dual descent (detailed in the sequel). Note that although Algorithm \ref{alg:iterative_pip_ob} does not include explicitly an update of the step size $a_{\ell}$ to  avoid notation cluttering, its subscript indicates it is iteration dependent.

\begin{algorithm}[t]
{\begin{small}
\caption{Iterative approximation for $\cP$ (IB relaxation - Pipage rounding)}
\label{alg:iterative_pip_ob}
\begin{algorithmic}
\State $\ell \gets 0$ 
\State $\lambda_{j\tau} \gets 0, \quad \forall j\in\cJ, \forall \tau\in\cT$
\While{$\ell < N$}
    \State $\vX^{(\ell)} \gets$ the solution of $\cP^{L{\text{(ib)}}}_{f_{LP}}(\vlambda)$ 
    \State $\vX^{(\ell)} \gets \texttt{Pipage}(\vX^{(\ell)})$
    \State $\lambda_{j\tau} \gets \lambda_{j\tau} - \alpha_{\ell} \left(c^{\text{(ib)}}_{j} - \sum_{i\in\cl I} {\mathbf{x}^{(\ell)}_{ij}}^\top \vp^{(\tau)}_{ij}\right), \quad \forall j\in\cJ, \forall \tau\in\cT$
    \State $\vX^{(\ell)} \gets \texttt{\textsf{Greedy-feasibility}}(\vX^{(\ell)})$
    \State $\ell\gets \ell+1$
\EndWhile
\end{algorithmic}
\end{small}}
\end{algorithm}

\subsection{Lagrangian Relaxation of OB Constraints}
\label{ssec:lag_ob}

Define $\cP^{L{(\text{ob})}}$ as $\cP$ with Lagrangian relaxation of the OB constraints, i.e.,
\begin{maxi}|l|[2]
{\vX, \vrho}
{{g(\vX) + \sum_{i\in\cl I}\sum_{t\in\cl T} \mu_{it}\left(c^{\text{(ob)}}_{i} - \sum_{j\in\cJ} x_{ijt}\right)} }
{\label{opt:lag_ob}}{\cP^{L{(\text{ob})}}: }
\addConstraint{\vX\in\cX^{\text{(ib)}}_{\text{int}},}{}
\end{maxi}
where $\vrho\in\bR_+^{I\times T}$ are the dual variables. Following a similar narrative as in Section \ref{ssec:lag_ib}, pipage is directly applicable to  $\cP^{L{(\text{ob})}}$ since the additional linear terms in the objective preserve the convexity of the pipage steps. Further, note that, as $\cP^{{(\text{ib})}}$ decouples $\forall j\in\cJ$, so does $\cP^{L{(\text{ob})}}$ for fixed $\vrho$. 

Define as $\cP_{f_{LP}}^{L{(\text{ob})}}(\vrho)$ the LP version of \eqref{opt:lag_ob} with fixed $\vrho$,
which can be also solved in parallel $\forall j\in\cJ$. Then, we can derive an iterative algorithm as follows: for fixed $\vrho$ obtain an initial solution via $\cP_{f_{LP}}^{L{(\text{ob})}}(\vrho)$ (which can be solved in parallel $\forall j \in \cJ$) and convert it to integral with pipage; then update $\vrho$ based on the updated integral solution. Repeat until convergence or a maximum number of iterations. In Algorithm \ref{alg:iterative_pip_ib} we summarize all steps of the iterative procedure.

% For the initial solution we can define the corresponding LP (for fixed $\vrho$) as
% \begin{maxi}|l|[2]
% {\vX}
% {{f(\vX) + \sum_{i\in\cl I}\sum_{t\in\cl T} \rho_{it}\left(c^{\text{(ob)}}_{i} - \sum_{j\in\cJ} x_{ijt}\right)} }
% {\label{opt:f_LPlag_ob}}{\cP_{f-LP}^{L{(\text{ob})}}(\vrho): }
% \addConstraint{\vX\in\cX^{\text{(ib)}},}{}
% \end{maxi}

\begin{algorithm}[t]
{\begin{small}
\caption{Iterative approximation for $\cP$ (OB relaxation - Pipage rounding)}\label{alg:iterative_pip_ib}
\begin{algorithmic}
\State $\ell \gets 0$ 
\State $\mu_{it} \gets 0, \quad \forall i\in\cI, \forall t\in\cT$
\While{$\ell < N$}
    \State $\vX^{(\ell)} \gets$ the solution of $\cP_{f_{LP}}^{L{\text{(ob)}}}(\vrho), \quad  / \star \text{ parallel } \forall j\in\cJ \star /$ 
    \State $\vX^{(\ell)} \gets \texttt{Pipage}(\vX^{(\ell)})$
    \State $\mu_{it} \gets \mu_{it} - \alpha_{\ell} \left(c^{\text{(ob)}}_{i} - \sum_{j\in\cl J} x_{ijt}^{(\ell)}\right), \quad \forall i\in\cI, \forall t\in\cT$
    \State $\vX^{(\ell)} \gets \texttt{Greedy-feasibility}(\vX^{(\ell)})$
    \State $\ell\gets \ell+1$
\EndWhile
\end{algorithmic}
\end{small}}
\end{algorithm}

As we saw in Section \ref{sec:ib}, solving in parallel $J$ ILPs (due to the decoupling) is fairly fast. We can leverage this fact here and propose an alternative version of Algorithm \ref{alg:iterative_pip_ib}. At each iteration, instead of solving $J$ LPs and then converting the solution to integral with pipage, we can solve directly the ILPs and obtain directly an integral solution. Define as $\cP_{f_{ILP}}^{L{(\text{ob})}}(\vrho)$ the ILP version of \eqref{opt:lag_ob} with fixed $\vrho$. Then, we can define a similar iterative procedure as in Algorithm \ref{alg:iterative_pip_ib} with the main difference being the way we update $\vX$. Algorithm \ref{alg:iterative_decouple} summarizes all steps.

%% file: evaluation.tex
\section{Numerical Evaluation}\label{sec:evaluation}

\subsection{Evaluation Setup, Datasets \& Metrics}

We employed the MIP solver FICO Xpress \citep{fico-xpress} for the direct ILP solutions, and the Gurobi solver \citep{gurobi} for the LP relaxations used in the pipage approximation algorithms. We ran the experiments in a c6a.\@12xlarge AWS EC2 instance with 48 vCPUs and 96GB memory, and we set a running time limit of $18k$ seconds for all solvers. This latter is particularly important as indeed in several cases the solvers hit this time limit.

Regarding the datasets, we used synthetic, yet realistic, networks and demand profiles of different types. In particular, we consider networks with increasing number of FCs, i.e., $I=10, 20, \dots, 50$, while the DSs are set to a fixed ratio to the FC number, i.e., $J/I =$ 1 and 2. These ratios and ranges correspond to diverse networks with a few hundred up to thousands of connections. We consider the general case with different cutoff times for each DS, drawn from a uniform distribution in a meaningful timeslot range, which, in conjunction with the varying transit times, yields intricate networks where the departure deadlines of each (potential) connection are spread throughout the day. Further, we set the number of product categories to $K =$ 250 and 500, where each category includes a randomly selected number of products in the range of $[100, 200]$. This product clustering is commonly used in e-commerce platforms to render their operations more tractable. Each product has its own expected demand profile, biased by the category it belongs to and the product itself. Further, we create a different demand profile for each DS so as to capture variations in ordering patterns across different regions and different hours of the day. Finally, for each network setting (values of $I$, $J$ and $K$) we generate 3 different instances and all results correspond to the mean value of these instances. For details regarding the dataset generation, including the network configuration and demand, we refer the reader to Appendix~\ref{app:data}.

The evaluation focuses on two key attributes of each algorithm. Firstly, its runtime or speed, i.e., how much time the algorithm needs to attain its result. Secondly, how this result compares with those returned by the benchmarks. Assessing this second attribute requires a careful approach. Namely, comparing directly the objective values attained by the algorithms (or their optimality gaps) can be misleading due to the problem structure: early demand can be easily covered by (almost) any algorithm that solves NDD and this inflates the objective value which, in turn, renders negligible (by comparison) the gains of an algorithm that succeeds in covering late demands. To alleviate this issue, we define the excess coverage as $g(\vX) - g_n$, where $g_n$ is the objective obtained by a naive algorithm. This metric focuses only on the additional demand a sophisticated algorithm can capture compared to that of a naive approach, and hence removes the impact of early demand that clutters these comparisons. To that end, we use a greedy algorithm (similar to the one in Section \ref{ssec:greedy_ob}) but selects the FC-DS connections in a random order, and deploys the last truck of each selected connection at the latest feasible time slot.

\begin{figure}[t]
	\centering
	\includegraphics[width=0.95\linewidth]{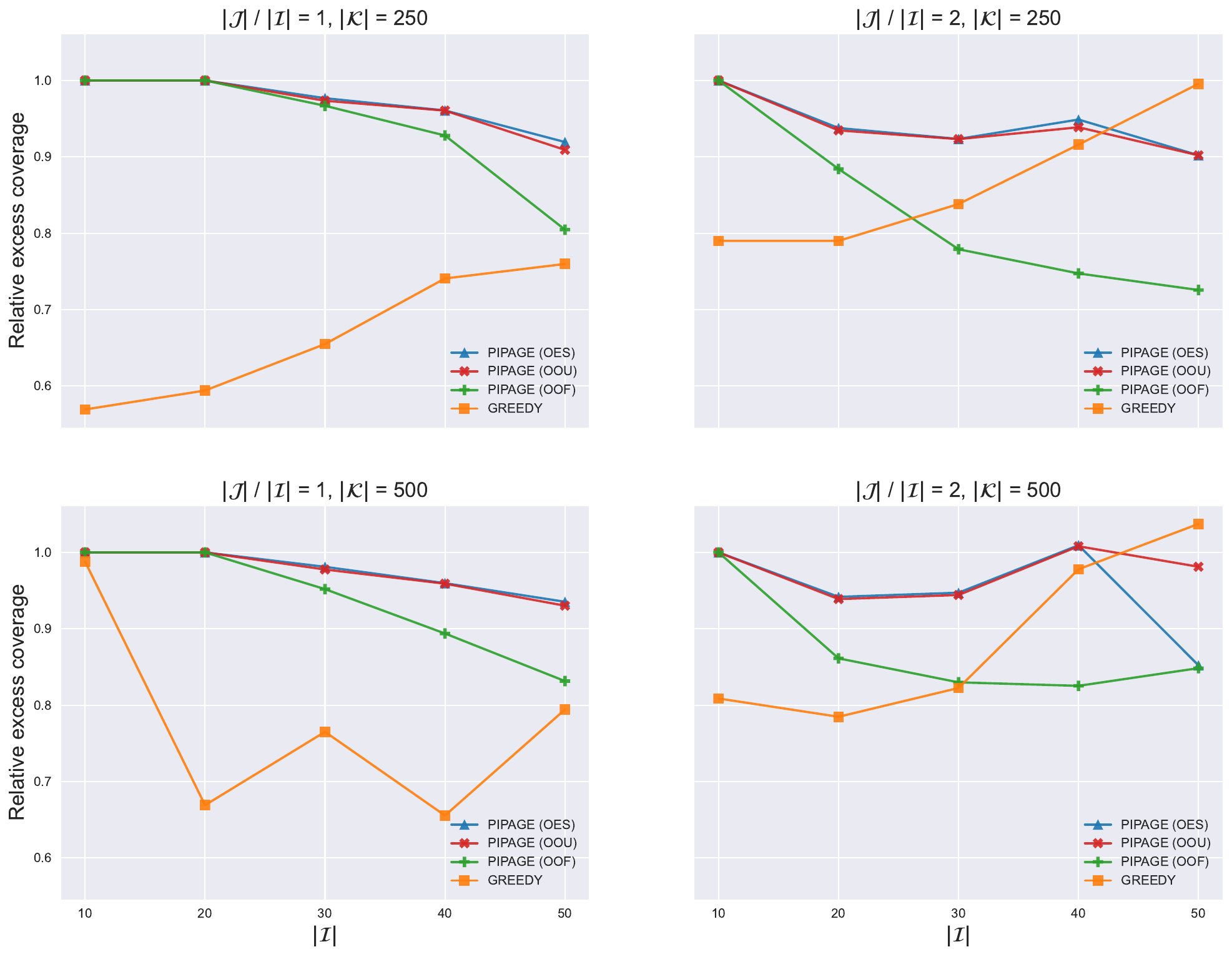}
	\caption{Relative excess coverage $E_{cov}(\vX)$ of all algorithms for $\cP^{\text{(ob)}}$. Optimality gap is measured with respect to the solution returned by a commercial MIP solver on a cloud server, with a time limit of 5 hours.}
	\label{fig:obj_ob}
\end{figure}

\begin{figure}[t]
	\centering
	\includegraphics[width=0.95\linewidth]{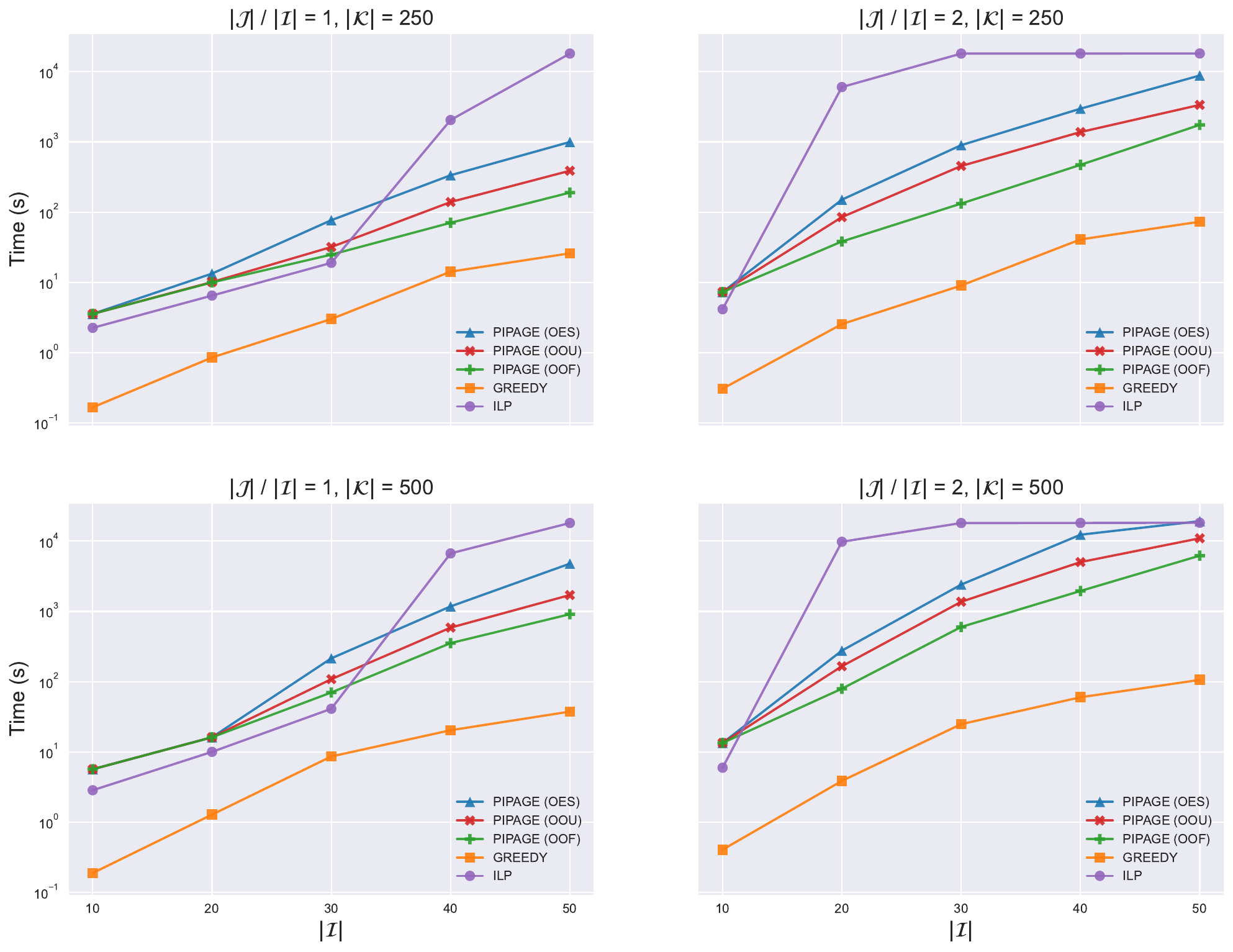}
	\caption{Running time of all algorithms for $\cP^{\text{(ob)}}$.}
	\label{fig:time_ob}
\end{figure}

\subsection{Evaluation of  $\cP^{\text{(ob)}}$} \label{ssec:ob_eval}

Figure~\ref{fig:obj_ob} presents the relative excess coverage of all algorithms for this problem, i.e., $E_{cov}(\vX)=(g(\vX) - g_n) / (g_{\text{ILP}} - g_n)$, for the various experiment configurations, where $g_{\text{ILP}}$ is the solution of the ILP formulation in Eq.\@ \eqref{opt:fx_ilp} by a commercial solver using the set time limit of 5 hours\footnote{The fact that a commercial solver running on well-provisioned served cannot find the optimal solution within this time limit, showcases exactly the computational challenge of the problem, and the need for faster solution strategies.}. When this ratio approaches the value of $1$, it means that the algorithm achieves the same objective as the ILP incumbent solution (not necessarily optimal due to the time limit). For smaller values, the ratio captures the \emph{relative optimality} gap and it is easy to verify that  $E_{cov}(\vX)\leq g(\vX)/g_{\text{ILP}}$, hence this is a tighter criterion indeed. Recall that functions $g(\cdot)$ and $f(\cdot)$ are equal for integer $\vX$, thus the ILP solution objective, $f(\vX^\star)$, is equal to $g(\vX^\star)\equiv g_{\text{ILP}}$. We observe that for smaller instances the greedy algorithm is clearly outperformed by all variations of the pipage approximation algorithm, which are very close to the solver's solution (relative optimal). For larger instances, the greedy algorithm's performance improves significantly and, in some settings, (mainly for $J/I = 2$) it even outperforms pipage. 

Figure~\ref{fig:time_ob} provides the running times. We observe that greedy is the fastest, by one or two orders of magnitude, while the pipage algorithms exhibit a trade-off between running time and objective, as discussed in Section \ref{sec:pip_strategies}. In particular, OES is marginally better than OOU and significantly better than OOF in terms of excess coverage, with the running time following the same pattern. The only exception is how the performance of OES degrades for $|\cI| = 50$, bottom right plot of Figure~\ref{fig:obj_ob}, which is due to the running time limit. As OES requires more time than the limit, it updates only few FCs and we use OOF for the remaining to allow the algorithm finish on time. It is interesting also to note that, unlike greedy and pipage, the solver time explodes for larger problems, reaching fast the adopted time limit. This is the reason that, in some plots, we see the greedy or pipage algorithm even achieve $E_{cov}(\vX)>1$. It is important to stress again that, for practical reasons, we have used a time limit each time we call the solver (independently or as part of an algorithm) to 18k seconds, or 5 hours. Hence, in order to understand how the solution time scales for each algorithm one needs to pay attention to the slope of the curves in Figure~\ref{fig:time_ob}; for instance, in upper right figure we see that the ILP hits the maximum time limit already for $I=30$.

Overall, there is no algorithm that outperforms all others in all settings. The quality and speed of solution is highly dependent on the particular network configuration and we can select the one that fits the needs of the application at hand, i.e., based on whether we prioritize the solution time or a higher objective is preferable. In terms of pipage variations, arguably the better trade-off is offered by OOU since its performance is marginally worse than OES, but exhibits a much better running time. Finally, we observe that the algorithms perform substantially better than their worst-case guarantees that were discussed in Section \ref{sec:theoretical}.

\subsection{Evaluation of $\cP^{\text{(ib)}}$}
\begin{figure}[t]
	\centering
	\includegraphics[width=0.95\linewidth]{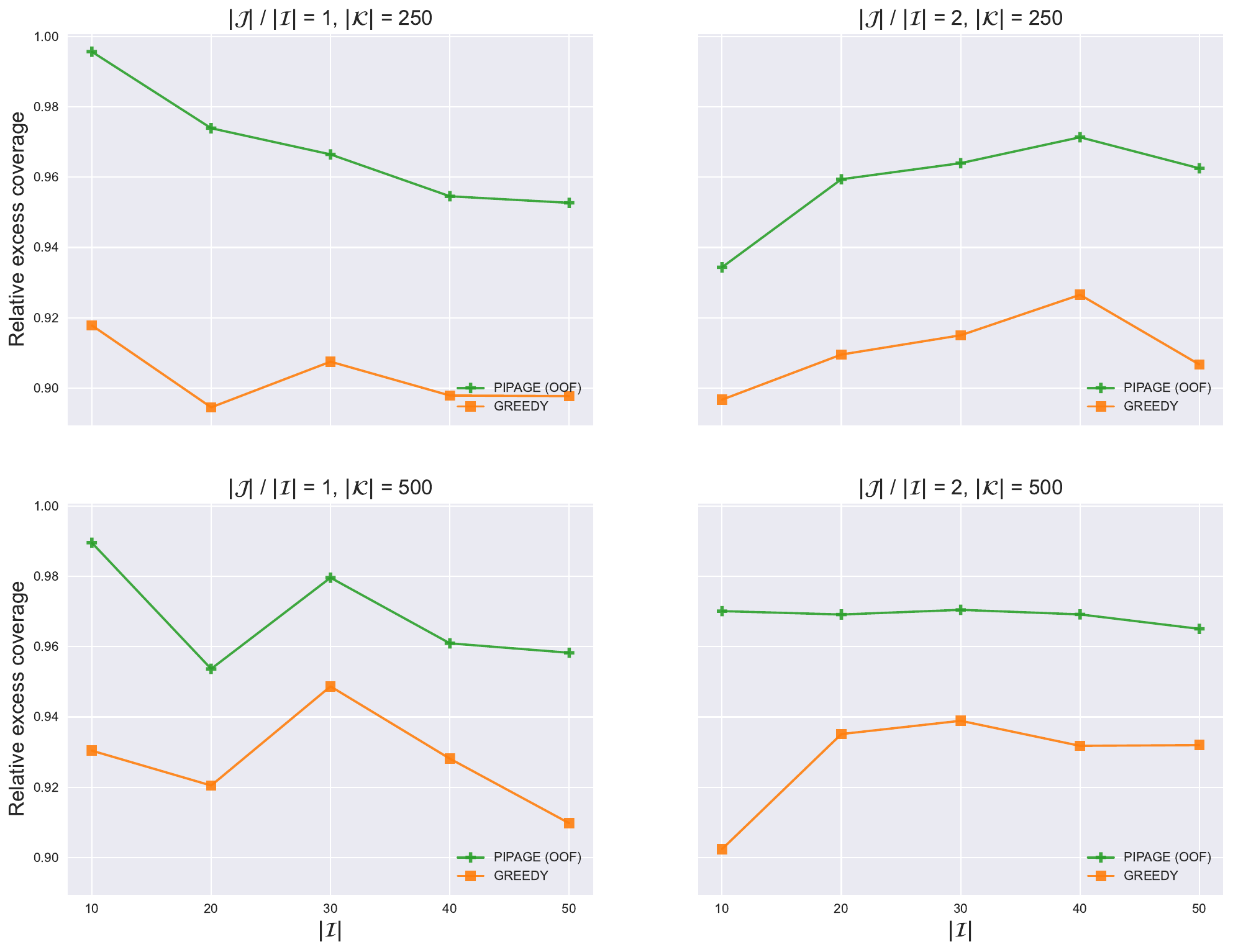}
	\caption{Relative excess coverage of all algorithms for $\cP^{\text{(ib)}}$.}
	\label{fig:obj_ib}
\end{figure}

\begin{figure}[h]
	\centering
	\includegraphics[width=0.95\linewidth]{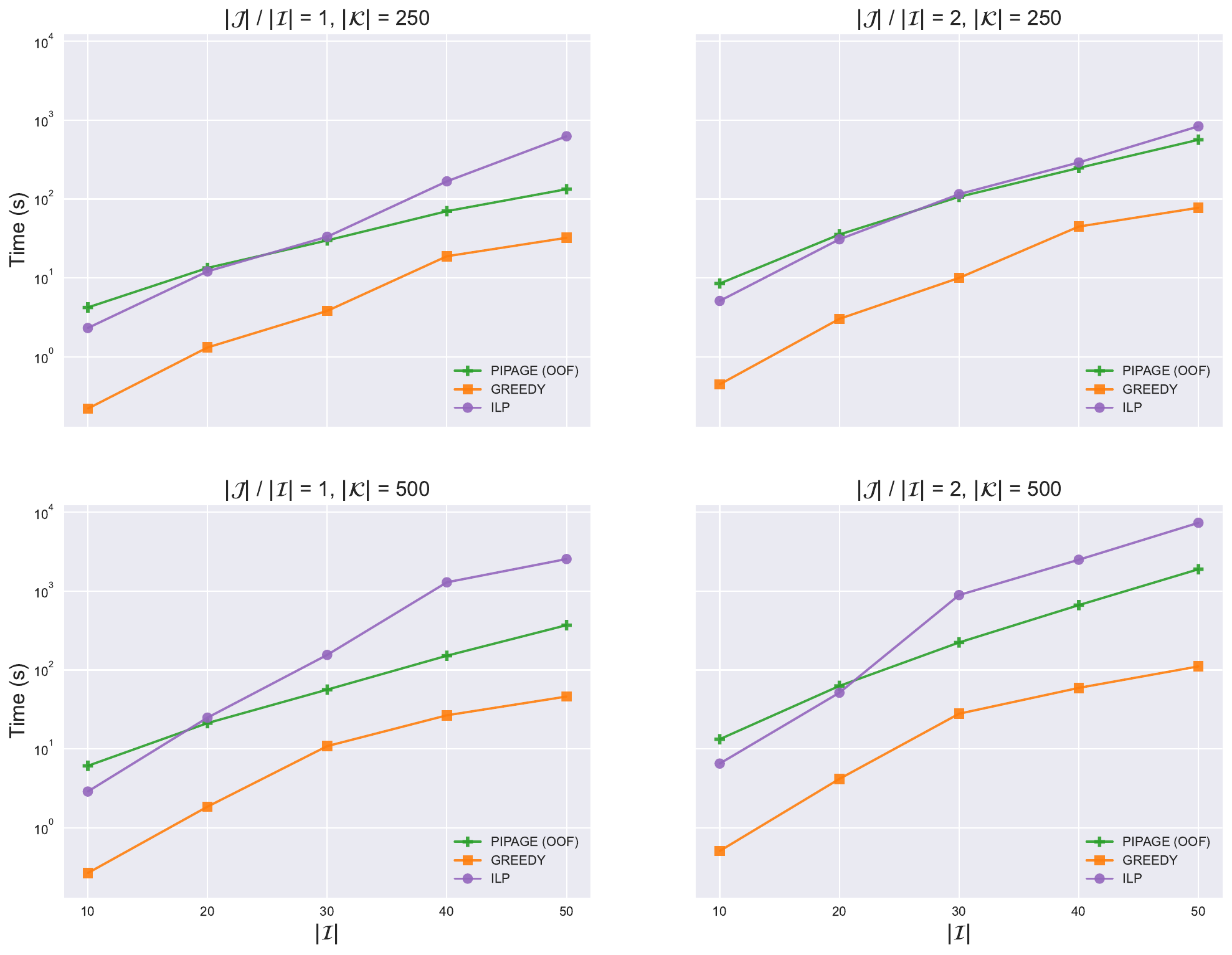}
	\caption{Running time of all algorithms for $\cP^{\text{(ib)}}$.}
	\label{fig:time_ib}
\end{figure}

Here, we compare the pipage solution and the benchmark greedy approach (as above) with an exact solution from a commercial solver that is subject to the time limit. In Figure~\ref{fig:obj_ib} we present again the relative (to a solver solution of the ILP formulation) excess coverage $E_{cov}$ of all algorithms, and compare their running times in Figure~\ref{fig:time_ib} (note the logarithmic scale). Due to the problem decomposition, the solver is significantly faster compared to the time it required for $\cP^{\text{(ob)}}$. As a result, it always finds the optimal solution within the time limit and its excess coverage is superior to both pipage and greedy by a margin of $1-5\%$ and $5-10\%$, respectively. Additionally, the solver's running time is similar to pipage for $|\cK| = 250$, while for $|\cK| = 500$ we see a margin for the bigger network sizes. 

We can see that there is a clear trade-off between speed and quality (attained objective value) for both approaches, which becomes more apparent for larger problem instances. Arguably, for small to medium networks, a solver can provide the optimal solution within reasonable time, which makes it an attractive option for $\cP^{\text{(ib)}}$. However, as the network size and the number of product categories grows, the approximation algorithms become more attractive in terms of running time. The selection of the algorithm depends, naturally, also on the opportunity (or, outsourcing) cost incurred by the e-retailer for each unit of uncovered demand. Finally, we observe again that both the greedy and pipage algorithms exceed their worst-case theoretical bounds, see Sec. \ref{sec:theoretical}, in these experiments. 

\subsection{Evaluation of $\cP$}
 
Finally, for this problem we compare all iterative methods from Sections \ref{ssec:lag_ib} and \ref{ssec:lag_ob} with the benchmark greedy approach, and a solution from a solver, as before. For the update of the dual variables we use the Polyak step \citep{polyak-step} which optimizes the descent at each iteration. Nevertheless, this step requires access to the (unknown) optimal function value. Here we resort to a lower bound of the optimal value, as we have a dual minimization problem.

\begin{algorithm}[t]
	\caption{Iterative approximation for $\cP$ (OB relaxation - Solver parallelization)}
 	\label{alg:iterative_decouple}
	\begin{algorithmic}
 		\State $\ell \gets 0$ 
 		\State $\mu_{it} \gets 0, \quad \forall i\in\cI, \forall t\in\cT$
 		\While{$\ell < N$}
 		\State $\vX^{(\ell)} \gets$ the solution of $\cP_{f_{ILP}}^{L{\text{(ob)}}}(\vrho), \quad / \star \text{ parallel } \forall j\in\cJ \star /$ 
 		\State $\mu_{it} \gets \mu_{it} - \alpha_{\ell} \left(c^{\text{(ob)}}_{i} - \sum_{j\in\cl J} x_{ijt}^{(\ell)}\right), \quad \forall i\in\cI, \forall t\in\cT$
 		\State $\vX^{(\ell)} \gets \texttt{Greedy-feasibility}(\vX^{(\ell)})$
 		\State $\ell\gets \ell+1$
 		\EndWhile
 	\end{algorithmic}
\end{algorithm}

In specific, we compute the dual step at each iteration $\ell$ using the formula:
\begin{align}
	\alpha_{\ell}=\frac{g_L^{(\ell)}-g_{feas}^{(\ell)}}{\|\vv^{(\ell)}\|_{2}^2}.
\end{align}
The denominator is the constraint violation error at the $\ell$-th iteration, i.e., $\vv^{(\ell)} = c^{\text{(ob)}}_{i} - \sum_{j\in\cl J} x_{ijt}^{(\ell)}$ or $\vv^{(\ell)} = c^{\text{(ib)}}_{j} - \sum_{i\in\cl I} {\mathbf{x}^{(\ell)}_{ij}}^\top \vp^{(\tau)}_{ij}$, depending on the iterative method, which constitutes the gradient for the dual descent. The numerator is an estimation of the current optimality gap of the dual problem. Namely, $g_L^{(\ell)}$ denotes the Lagrangian, i.e., the objective of pipage (or the solver, if used) at the $\ell$-th iteration (before the feasibility correction) which matches the dual function value $h(\mathbf \lambda_{\ell})$ since, for the IB-relaxed problem, it holds:
 \[
 h(\mathbf{\lambda}_{\ell})=\max_{\vX\in\cX^{\text{(ob)}}_{\text{int}}} \ \left\{ {g(\vX) + \sum_{j\in\cl J}\sum_{\tau\in\cl T} \lambda_{j\tau}\left(c^{\text{(ib)}}_{j} - \sum_{i\in\cl I} \mathbf{x}_{ij}^\top \vp^{(\tau)}_{ij}\right)} \right\}.
 \] 
On the other hand, $g_{feas}^{(\ell)}$ is calculated after maximizing the Lagrangian and correcting the obtained solution for capacity feasibility (recovering the relaxed constraint). This is clearly smaller than the current optimal value of $g$ and hence can serve as a lower bound for the (unknown) optimal value of the dual function $h(\mathbf{\lambda}^*)$. We can devise in a similar fashion the dual step for the OB-relaxed problem.

 \begin{figure}[t]
 	\centering
 	\includegraphics[width=0.95\linewidth]{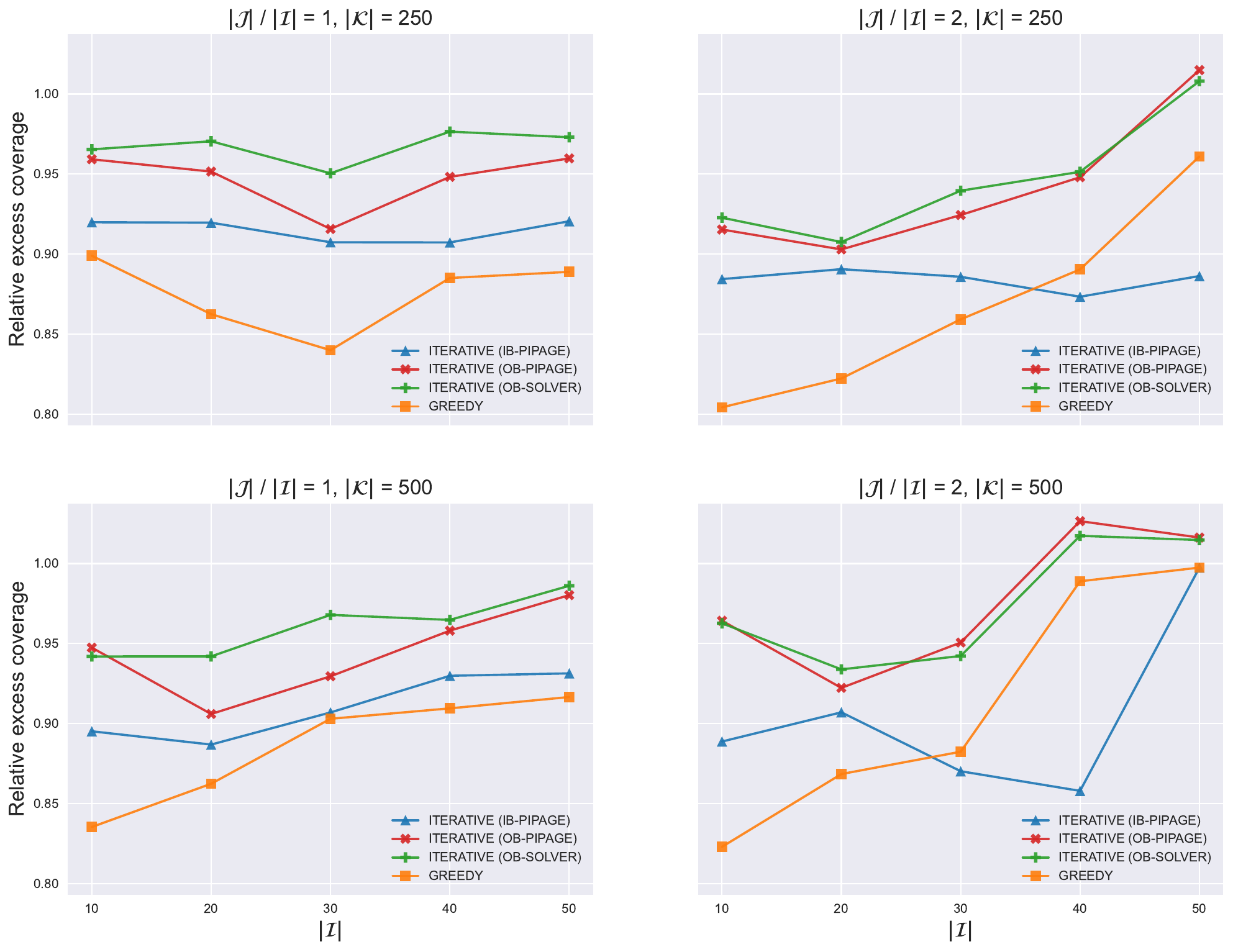}
 	\caption{Relative excess coverage of all algorithms for $\cP$.}
 	\label{fig:obj_iob}
 \end{figure}
 
 \begin{figure}[t]
 	\centering
 	\includegraphics[width=0.95\linewidth]{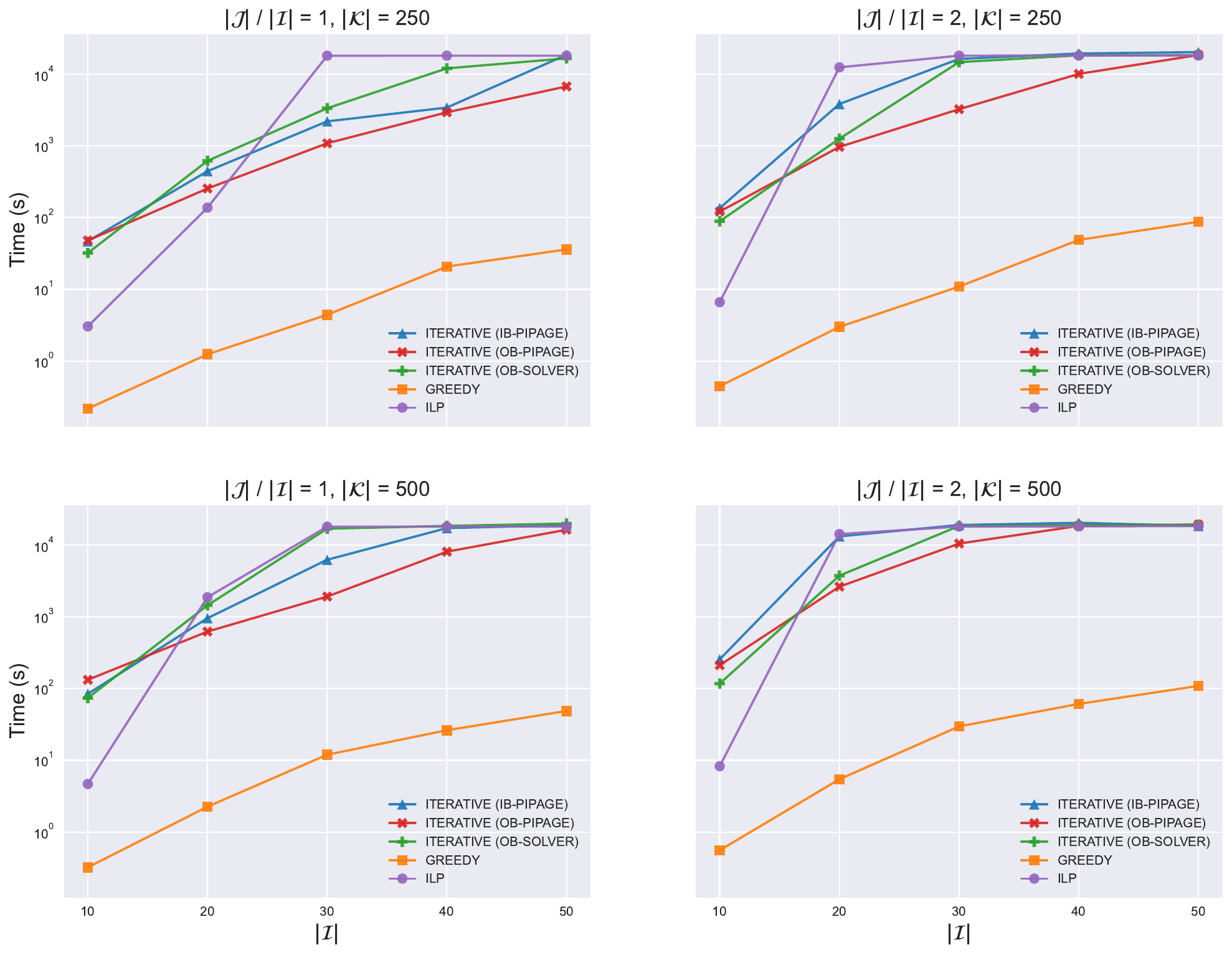}
 	\caption{Running time of all algorithms for $\cP$ on a logarithmic scale, for different network sizes.}
 	\label{fig:time_iob}
 \end{figure}

Finally, we set the maximum number of iterations to $N=100$, with 20 iterations \emph{patience}, i.e., the algorithm terminates if there is no improvement for 20 consecutive iterations (also not explicitly included in Algorithms \ref{alg:iterative_pip_ob}-\ref{alg:iterative_decouple} to avoid cluttering), and we keep the best solution out of all iterations. Other typical termination criteria one can use include assessing the numerator of the step $\alpha_{\ell}$, or the distance from the solution of the relaxed problem, and so on. We found the first termination criterion sufficient and well-performing for our problems. 
 
Our findings suggest that the iterative methods outperform the greedy algorithm in terms of coverage with a 5-10\% margin, while the running time of the greedy is much lower, as expected. Among the iterative methods, the two OB relaxation methods (Algorithms \ref{alg:iterative_pip_ib} and \ref{alg:iterative_decouple}) clearly outperform the IB relaxation (Algorithm \ref{alg:iterative_pip_ob}), since they achieve much better coverage with similar or even smaller running time. This is due to the decoupling property of these algorithms, which results in faster and more accurate iterations, and therefore the algorithms converge faster to better solutions. This is more pronounced in the larger settings, where the coverage of Algorithm \ref{alg:iterative_pip_ob} is significantly lower, since only few iterations are possible within the adopted time limits. Further, there is a natural trade-off between Algorithms \ref{alg:iterative_pip_ib} and \ref{alg:iterative_decouple} in terms of coverage and running time. In general, solver iterations are more accurate but slower, while pipage iterations are faster but less accurate, which is evident in Figures \ref{fig:obj_iob}  and \ref{fig:time_iob}. Finally, it is important to note that for the larger networks where both the ILP solution and the iterative algorithms consume all the available running time, Algorithms \ref{alg:iterative_pip_ib} and \ref{alg:iterative_decouple} outperform the solver, i.e., they converge faster to a better solution.

%% file: discussion.tex
\section{Discussion \& Conclusions}\label{sec:discussion}

This paper proposed a new problem related to middle-mile next-day delivery (NDD) that has important operation implications, yet is hitherto overlooked. Namely, we introduce the problem of last-truck scheduling for NDD coverage, and explain why this middle-mile design aspect is crucial from an operations point of view, and how it differs from the typical truck dispatching problem or the network design problem. Next, we propose an optimization model that selects the time and destination of last trucks departing from fulfillment centers towards a network of delivery stations. The obtained schedule determines the range of products and orders that can be served (or, \emph{covered}) with next-day delivery promise. We consider different variants of the problem, depending on the operational constraints that different e-retailers might have, and characterize their properties and complexity. As our own experiments showcase, even small to moderate problem instances are challenging to solve in practice (i.e., within few hours) with state-of-the-art commercial MIP solvers. Hence, we create an optimization toolbox for their solution, using a greedy algorithm, pipage-based approximation algorithms, and Lagrangian heuristics for the most constrained problem variant; all of which expedite the solution by orders of magnitude compared to solvers. Our extensive experiments highlight an interesting trade-off between scalability and performance of these solution approaches, and one can select the most suitable approach for the problem at hand. In all cases, using realistic and diverse scenarios, we found that the performance of the best approximation algorithm is no less than 90\% of the NDD coverage solution that is found by commercial MIP solvers, while reducing the solution time by orders of magnitude. 

We believe this work paves the road for several interesting next steps. For instance, one can naturally expand the model and solution methods to optimize more than one last trucks per time period (e.g., two per day), if this is necessary due to the operational requirements of some retailer. Similarly, while direct lanes for last trucks (as we consider here) are the most natural choice, consolidation savings might be imperative in some other cases (e.g., small-scale retailers). Thus, extending this approach to multihop and multipath networks, possibly with consolidation capacities and delays, is another non-trivial extension of practical importance. In terms of solution methods, it remains an open problem to design a computationally-efficient algorithm with non-trivial optimality guarantees (better than greedy) for the problem with both inbound and outbound constraints; and it is of further interest to explore combination of the presented techniques, e.g., using the greedy algorithm as a warm-start strategy for a MIP solver.

%% file: appendix.tex
\section{APPENDIX}

\section{Pipage Step}
\label{sec:pipage_step}

The general idea of a pipage step is the following: given any non-integral solution $\vX$, find a direction $\vZ$ and scalars $\epsilon^+, \epsilon^- > 0$, such that $\vX + \epsilon^+\vZ$ and $\vX - \epsilon^-\vZ$ are feasible (apart from the integrality constraint) and have strictly more integral coordinates than $\vX$. Further, the objective $g(\cdot)$ should be convex in this direction, i.e., $g(\vX + \gamma\vZ)$ is a convex function over variable $\gamma$. If these conditions are met, it is ensured that with every pipage step we convert at least one non-integral variable to integral, and the objective does not decrease (due to convexity). Unfortunately, there is no standardized way to find such a direction $\vZ$, hence in the following we describe a tailored approach for finding such directions for each NDD variant.

\subsection{Pipage Step for $\cP^{\text{(ob)}}$}
\label{sec:pipage_step_ob}

First, let us examine the set of constraints that a pipage step needs to satisfy in order to remain feasible. For $\cP^{(\text{ob})}$, apart from constraint \eqref{opt:basic_mip_b} that sets to $0$ all timeslots after the departure deadline, we need to satisfy constraints \eqref{opt:basic_mip_c} (at most one last truck) and \eqref{opt:basic_mip_d} (OB truck constraints), which we rewrite here for convenience:
\begin{align*}
\mathbf{x}_{ij}^\top \mathbf{1}_T \leq 1, & \quad\forall i \in \cl I, \ \forall j\in\cl J, \\
\sum_{j\in\cl J} x_{ijt} \leq c^{\text{(ob)}}_{i}, & \quad \forall i \in \cl I, \ \forall t\in\cT.
\end{align*}
These constraints should be satisfied $\forall i \in \cI$, i.e., the constraints decouple for each FC. This implies that a feasible pipage step which applies changes only to one FC is ensured to retain feasibility across all FCs. Therefore, we focus on finding directions for the pipage step that apply feasible updates for a fixed FC $i\in\cI$. 

The above formulation (given a fixed $i\in\cI$) falls under the standard pipage problems with two-dimensional step updates, and boils down to finding cycles or paths whose endpoints have degree 1 and applying symmetric changes \citep{ageev2004pipage}. Although it is possible to find directions that satisfy both constraints, here we leverage the fact that \eqref{opt:basic_mip_c} is not essential and we can drop it. This simplifies significantly the problem since now we need to search for directions in one dimension, with fixed $i\in\cI$ and $t\in\cT$. 

\begin{algorithm}[t]
{\begin{small}
\caption{Pipage step for $\cP^{\text{(ob)}}$}\label{alg:step}
\begin{algorithmic}
\For{$i \gets 1$ to $I$}
\For{$t \gets 1$ to $T$}
    \State $n \gets$ number of non-integer elements of $\vx_{it}$
    \While{$n > 1$}
        \State $j_1, j_2 \gets$ indexes of two non-integer entries of $\vx_{it}$
%        \State $\vZ \gets \vzero$, $x_{ie_1t} \gets 1$, $x_{ie_2t} \gets -1$
        \State $\vZ \gets \vzero$, $z_{i j_1, t} \gets 1$, $z_{i j_2, t}  \gets -1$        
        
        \State $\epsilon^+ = \min(1 - x_{ij_1t}, x_{ij_2t})$, $\epsilon^- = \min(1 - x_{ij_2t}, x_{ij_1t})$
        \If{$g(\vX + \epsilon^+ \vZ) \geq g(\vX)$}
            \State $\vX \gets \vX + \epsilon^+ \vZ$
        \Else
            \State $\vX \gets \vX - \epsilon^- \vZ$
        \EndIf
        \State $n \gets$ number of non-integer elements of $\vx_{it}$
    \EndWhile
\EndFor
\EndFor
\end{algorithmic}
\end{small}}
\end{algorithm}

Finding pipage directions in a vector is detailed in \cite{ageev2004pipage}. For completeness, we give a high level overview of this process in Algorithm \ref{alg:step}. In a nutshell, for every vector $\vx_{it} \in [0, 1]^J$ we select randomly  any two non-integer entries, say with indices $j_1$ and $j_2$. The pipage step direction is defined by vector $\vz$, where $z_j = 0, \forall j\in \cJ /\{j_1, j_2\}$, $z_{j_1} = 1$ and $z_{j_2} = -1$. In the general case (ignoring here the sign and magnitude) we apply a symmetric change $\epsilon$ to $\vx$ towards the direction $\vz$, so it holds:
\begin{equation*}
    (\vx_{it} + \epsilon\vz)^\top\vone_J = \sum_{j\in \cJ /\{j_1, j_2\}} x_{ijt} + (x_{ij_1t} + \epsilon) + (x_{ij_2t} - \epsilon) = \vx_{it}^\top\vone_J \leq c^{\text{(ob)}}_{i},
\end{equation*}
i.e., the updated vector satisfies \eqref{opt:basic_mip_d}. By construction, in each iteration at least one element becomes integer. Further, the objective does not decrease since the direction is convex in terms of $\epsilon$ as $g(\vX + \epsilon\vZ) = \alpha\epsilon^2 + \beta \epsilon + \gamma$, with $\alpha \geq 0$. For clarity of exposition, we did not include the case where only one element is non-integer; this is trivial and the entry can be updated to 1 or 0 by comparing the corresponding objectives. Also, since we dropped \eqref{opt:basic_mip_c}, pipage might produce solutions with multiple trucks for some FC-DS pairs. As discussed, such solutions can be remedied by dropping all earlier trucks. Finally, it is important to highlight that Algorithm \ref{alg:step} can run in parallel for all $i\in\cI$.

\subsection{Pipage Step for $\cP^{\text{(ib)}}$}
\label{sec:pipage_step_ib}

In a similar manner, in the $\cP^{(\text{ib})}$ problem the set of constraints that a pipage step needs to satisfy  are \eqref{opt:basic_mip_c} (at most one last truck) and \eqref{opt:basic_mip_e} (IB truck constraints), i.e., 
\begin{align*}
\mathbf{x}_{ij}^\top \mathbf{1}_T \leq 1, & \quad\forall i \in \cI, \ \forall j\in\cl J, \\
\sum_{i \in \cl I} \vx_{ij}^\top \vp^{(\tau)}_{ij} \leq c^{\text{(ib)}}_{j}, & \quad \forall j\in \cJ, \ \forall\tau\in \cl T.    
\end{align*}
The constraints should be satisfied $\forall j \in \cJ$, i.e., they decouple for each DS. This implies that a feasible pipage step which applies changes only to one DS retains feasibility across all DSs. Therefore, we focus on finding directions for the pipage step that apply feasible updates for a fixed DS $j\in\cJ$. 

Unlike the case with OB constraints presented in Appendix \ref{sec:pipage_step_ob}, to the best of our knowledge there is no systematic way of finding convex directions that satisfy both \eqref{opt:basic_mip_c} and \eqref{opt:basic_mip_e} since a generic path or cycle direction $\vZ$ results in $g(\vX + \epsilon\vZ) = \alpha\epsilon^p + \beta\epsilon^{p-1} + \dots + \gamma$, with $p > 2$. This is due to the nature of the objective. For fixed $i\in\cI$ (as in Appendix \ref{sec:pipage_step_ob}), a path or cycle over $\vX_i\in \{0, 1\}^{J\times T}$ results in an objective $g(\vX + \epsilon\vZ)$ with additive (due to summation over $j$) quadratic terms (due to product over $t$). However, for fixed $j\in\cJ$, a path or cycle over $\vX_j\in \{0, 1\}^{I\times T}$ results in multiplicative (due to product over $i$) quadratic terms (due to product over $t$).

As it becomes obvious, the fact that we can drop \eqref{opt:basic_mip_c}, as in the OB case,  becomes essential for the application of pipage in $\cP^{\text{(ib)}}$. Again, we need to search for directions in 1D with fixed $j\in\cJ$ and $\tau\in\cT$. The procedure is similar to Algorithm \ref{alg:step} and therefore omitted.

\section{Proofs}\label{sec:proofs}

\subsection{Proof of Theorem \ref{the:complexity}}

\subsubsection{$\cP^{\text{(ob)}}$ is NP-hard.}
\label{appendix:reduction_ob}

For this proof, we use a reduction from the femtocaching problem \citep{femtocaching}, which in turn uses a reduction from the 2-Disjoint set cover problem. We start with the description of the caching problem, and in order to simplify the presentation, we focus directly on the optimization problem (not the decision problem). We are given a set $\mathcal H$ of caches (i.e., helpers) that can store up to $M$ files; a set $\mathcal F$ of files of equal size; and a set of users $\mathcal U$, where each user $u\in\mathcal U$ is connected to a subset of the caches $\mathcal H(u)$. We denote with $p_f, f\in\mathcal F$ the probability of a request for file $f$, and with $w_u$ the gain (e.g., in transmission delay) when user $u$ retrieves a file from some cache. The problem asks to decide which files to store at each cache in order to maximize the caching benefits across all users. This can be expressed as follows:
\begin{maxi*}|l|[2]
{\vx}
{\sum_{f\in\mathcal F} \sum_{u\in\mathcal U}p_fw_u \left(1-\prod_{h\in\mathcal H(u)} \left(1-x_{fh}\right)\right)}{}{}
\addConstraint{\sum_{f\in\mathcal F}x_{fh}\leq M, \ \ \forall h\in\mathcal H.}
\end{maxi*}
Now, given an instance of femtocaching, we can create in polynomial time an instance of $\cP^{\text{(ob)}}$ as follows. We set $T=1$ and create the set of DSs $\mathcal J$ based on the set of files $\mathcal F$ (one DS per file); the set of products $\mathcal K$ based on the set of users $\mathcal U$ (one product per user); and the set of FCs $\mathcal I$ based on the set of helpers $\mathcal H$ (one FC per helper). We also set $c_i^{(ob)}=M$ for each FC $i$. Then, we can rewrite $\cP^{\text{(ob)}}$ as
\begin{maxi*}|l|[2]
{\vx}
{\sum_{j\in\mathcal J} \sum_{k\in\mathcal K}d_{jk} \left(1-\prod_{i\in\mathcal I(k)} \left(1-x_{ij}\right)\right)}{}{}
\addConstraint{\sum_{j\in\mathcal J}x_{ij}\leq M, \quad \forall i\in\cI,}
\end{maxi*}
where $\mathcal I(k)\subseteq \mathcal I$ is the subset of FCs that store item $k\in\mathcal K$. It becomes clear that if we could solve this problem in polynomial time, then we could also solve the above femtocaching problem.

\subsubsection{$\cP^{\text{(ib)}}$ is Inapproximable}
\label{appendix:reduction_ib}

We show that $\cP^{\text{(ib)}}$ is NP-hard to approximate within $1-1/e$ of its optimal solution using a reduction from the maximum coverage problem with cardinality constraints (Max-k-coverage). We first introduce this latter problem. Given an integer number $k<m$, a universe of base elements $\mathcal U=\{u_j\}_{j=1,\dots, n}$, and a set of (possibly overlapping) sets $\mathcal L=\{\cS_1, \dots, \cS_m\}$ with $\cS_i\subset \cU, \forall i\leq m$, we wish to choose at most $k$ sets from $\mathcal L$, so as to maximize the covered elements of their union. 
% $|\cup_{i\!=\!1}^k\mathcal{S}_{i}^{'}|$. 
If we introduce the variable $x_i\in \{0,1\}$ with $x_i=1$ if $\cS_i$ is selected, and $y_j\in \{0,1\}$, with $y_j=1$ if element $u_j$ is covered, we can express this problem as a MIP:
\begin{maxi*}|l|[2]
{\vx, \vy}
{\sum_j y_j \qquad \qquad \qquad\qquad \;\; \text{(maximize coverage)}}
{}{}
\addConstraint{\sum_i x_i\leq k,}{}{\quad \text{(choose at most $k$ sets)}}{}
\addConstraint{\sum_{i:\cS_i\ni e_j} x_i \geq y_j, \quad \forall j,}{}{\quad \text{(element coverage condition)}}{}
\addConstraint{\vx\in \{0,1\}^m, \vy\in\{0,1\}^n.}{}
\end{maxi*}
In 1978, \cite{nemhauser1978analysis} showed that there is no polynomial-time algorithm that can provide a solution  $\geq (1-1/e) \cdot  \texttt{opt} \approx 0.632 \cdot \texttt{opt}$ unless P $=$ NP, and that a greedy algorithm that constructs a solution by progressively adding the set that provides the largest differential coverage achieves the optimal approximation.

It can be easily seen that the above problem is polynomial-time reducible to a certain instance of $\cP^{\text{(ib)}}$. Indeed, given an instance of Max-k-coverage, we can set $T=1$, create the set of demands $\mathcal D$ by introducing a unit demand element for each element in $\mathcal U$; create $m$ FCs, where each FC $i$ may dispatch a truck that covers the demand elements in the respective set $\cS_i$ from $\mathcal L$, and create 1 DS with inbound capacity $k$. Then, solving this instance of $\cP^{\text{(ib)}}$, would give us the solution for this, general, max-k-coverage problem. Furthermore, it follows that if we could approximate the optimal solution of this instance within $1-1/e$, this would induce a similar approximation for the Max-k-coverage problem.

\subsection{Proof of Lemma \ref{lem:submodular}} \label{appendix:proof-submodular-function}

Let us start by considering two sets $\cl S$ and $\cl {\tilde S}$ with $\cl S \subset \cl {\tilde S}$, and let us denote with $\vx_{ij}$ and $\tilde{\vx}_{ij}$, $\forall i,j$, the elements of the respective assignment matrices, $\vX$ and $\tilde{\vX}$. Then, it follows that
\[
\sum_{j\in\cJ, k\in\cK, t\in\cT}
d_{jkt}\left( 1 - \prod_{i\in\cI, \tau\geq t} \left(1 - a_{ik} x_{ij\tau}\right)\right) \leq 
\sum_{j\in\cJ, k\in\cK, t\in\cT}
d_{jkt}\left( 1 - \prod_{i\in\cI, \tau\geq t} \left(1 - a_{ik} \tilde{x}_{ij\tau}\right)\right)
\]
since it holds
\[
\prod_{i\in\cI, \tau\geq t} \left(1 - a_{ik} x_{ij\tau}\right) \geq \prod_{i\in\cI, \tau\geq t} \left(1 - a_{ik} \tilde{x}_{ij\tau}\right), \quad \forall j\in\cJ.
\]
Indeed, if the demand at some DS $j\in\cJ$ is not covered by $\tilde{\vx}_{ij}$ then it is certainly not covered by $\vx_{ij}$, as $\vX$ schedules a subset of trucks of $\tilde{\vX}$. In other words, adding more trucks in $\vX$ in order to create the larger schedule $\tilde{\cS}$ can only increase the covered demand. Hence, $g(\cS)$ is monotone increasing on $\cS$.

Regarding submodularity, since the sum of submodular functions is also a submodular function, we will focus on the covered demand at a specific DS $j\in\cJ$:
\[
g_j(\cS)=\sum_{k\in\cK, t\in\cT}d_{jkt}\left( 1 - \prod_{i\in\cI, \tau\geq t}\left(1 - a_{ik} x_{ij\tau}\right)\right).
\]
We consider two schedules $\cS \subset \tilde{\cS}$, and a new scheduled truck on link $(i,j)$ at slot $t$, that is represented with the element $\hat v_{ijt}$, where $\hat v_{ijt}\notin \tilde{\cS}$ (which also implies $\hat v_{ijt}\notin \cS$). We evaluate the increase in the value of $g_j$ for the following cases that capture all possible scenarios when we add this new truck on either schedule. Namely, we consider the following cases:
\begin{enumerate}
    \item If $\exists t'\in\cT$, with $t'>t$, such that $v_{ijt'} \in \cS$ (and hence $v_{ijt'} \in \tilde{\cS}$), then:
    \[
        g_j\left(\tilde{\cS} \cup \{\hat v_{ijt}\}\right)-g_j\left(\tilde{\cS}\right) = g_j\left(\cS \cup \{\hat v_{ijt}\}\right) - g_j\left(\cS\right) = 0,
    \]
    since the demand that the new truck $\hat v_{ijt}$ covers was already covered by the later truck $v_{ijt'}$.
    \item If $\exists t'\in\cT$, with $t'>t$, such that $v_{ijt'} \in \tilde{\cS}$ but $v_{ijt'} \notin \cS$, with $t'>t$, then:
    \[
        g_j\left(\tilde{\cS} \cup \{\hat v_{ijt}\}\right) - g_j\left(\tilde{\cS}\right) = 0 \qquad \text{and} \qquad g_j\left(\cS \cup \{\hat v_{ijt}\}\right) - g_j\left(\cS\right) \geq 0,
    \]
    since truck $\hat v_{ijt}$ covers potentially uncovered demand in $\cS$, but not in $\tilde{\cS}$. 

    \item If $\exists t'$, with $t'<t$, such that $v_{ijt'}$ is the latest truck scheduled in $(i,j)$ under $\cS$, then it also appears in $\tilde{\cS}$, which might have  additionally later trucks on that connection. Hence:
        \[
            g_j\left(\cS \cup \{\hat v_{ijt}\}\right) - g_j\left(\cS\right)\geq g_j\left(\tilde{\cS} \cup \{\hat v_{ijt}\}\right) - g_j\left(\tilde{\cS}\right) \geq 0,
        \]
    since truck $\hat v_{ijt}$ might cover demand in $\cS$ than is (partially) covered in $\tilde{\cS}$.
\end{enumerate}
Hence, the property of submodularity (diminishing returns) is satisfied in any case.

\subsection{Proof of Lemma \ref{lem:partition-intersection-all}}
\label{appendix:proof-matroid}

Let us start by providing the definition of partition matroid. We denote with $\cl E$ the finite ground set (i.e., set of all elements) and assume we can express this set as the union of $N$ disjoint subsets (or, partitions) $\cl E_1, \cl E_2, \ldots, \cl E_N$. We are also given the numbers $k_1, k_2, \ldots, k_N$. Then, we define the family $\cl Q$ of independent sets of this matroid as:
\begin{align*}	
	\cl Q=\Big\{ \cl S\subseteq \cl E \, : \, \big | \cl S \cap \cl E_n \big | \leq k_n, \, \forall n=1,\ldots, N	 \Big\}
\end{align*}
Next, we define the partition matroid $\cl M=(\cl E, \cl Q)$, with ground set $\cl E$ and independent sets drawn from $\cl Q$. 
Using this definition, we can create one partition matroid for each type of constraints in $\cP$, and then express the overall constraints as the intersection of these matroids. In detail, we define the set of ground elements:
\begin{align*}
	\cl V=\{ v_{ijt}\}_{i\in \cl I, j\in \cJ, t\in \cT},
\end{align*}
where $v_{ijt}$ represents the decision to send a truck from FC $i\in\cI$ to DS $j\in\cJ$ at slot $t\in\cT$. Using this set, we can define two different families of independent sets. The first one corresponds to \eqref{opt:basic_mip_d} and is: 
\begin{align*}
\cl Q_1 = \{\cS \subseteq \cl V : | \cl S \cap \cl V_{it} | \leq c_i^{(ob)}, \,\, \forall i \in \cl I, t\in \cl T\},
\end{align*}
where $\cl V_{it} =\{ v_{i1t}, v_{i2t}, \dots, v_{iJt} \}, \forall i\in \cI, t\in \cT$, are the partitions of the ground set for this second matroid. For the constraint \eqref{opt:basic_mip_e} we define the matroid:
\begin{align*}
	\cl Q_2 = \{\cS \subseteq \cl V : | \cS \cap \cl V_{jt} | \leq c_j^{(ib)}, \,\, \forall j \in \cJ, t\in \cT\},
\end{align*}
where $\cl V_{jt} = \{ v_{1jt}, v_{2jt}, \dots, v_{Ijt} \}, \forall j\in \cJ, t\in \cT$, are the partitions for this second matroid.
Every feasible truck schedule corresponds to an element of matroid $M=(\cl V, \cl Q)$ where $\cl V$ is the set of all possible trucks and $\cl Q=\cl Q_1 \cap \cl Q_2$ is the family of sets of edges that satisfy concurrently all constraints, i.e., are independent in the two matroids. Similarly, we can define the constraints of problems $\cP^{\text{(ib)}}$ and $\cP^{\text{(ob)}}$ using only the respective partition matroid. 

For the second part of the lemma, it suffices to show with a counter-example that a defining matroid property is not satisfied. Indeed, a simple example reveals that the \emph{augmentation property} does not hold. Assume we have a network with 2 FCs and 4 DSs operating for $T=2$ timeslots, where the inbound capacity of all DSs are equal to $c_j^{(ib)}=1$, for $j=1,\dots, 4$, and for the FCs it holds $c_1^{(ob)}=2$ and $c_2^{(ob)}=1$. Consider now the following two schedules:
\begin{align*}
&\mathcal S_1=\{ (i,j,t): (1,2,1), (1,3,1), (1,1,2), (1,4,2), (2,3,1), (2,4,1), (2,3,2) \}, \\
&\mathcal S_2=\{ (i,j,t): (1,1,1), (1,3,1), (1,2,2), (1,4,2), (2,3,2), (2,4,1) \}, 
\end{align*}
where each of them is independent, i.e., satisfies the inbound and outbound capacities. Also, it holds $|\cS_1| > |\cS_2|$ ($7>6$) with $\mathcal{S}_1\setminus \mathcal{S}_2=\{ (1,2,1), (1,1,2), (2,3,1)\}$. However, none of these two elements can be added to $\mathcal S_2$ to yield an independent (feasible) schedule: adding $(1,2,1)$ and $(1,1,2)$ would violate FC $1$'s outbound constraint; while adding $(2,3,1)$ would violate the inbound constraint of DS $3$.

\subsection{Proof of Lemma~\ref{lemma:rho}}
\label{app:proof_lemma_rho}

We want to prove that $g(\vX) \geq \rho(mT) \cdot f(\vX)$, where $\rho(x)= 1 - (1 - 1/x)^x$ and $m$ the maximum number of inbound connections to any destination node  $j\in\cJ$.

Fix $j, t, k$ and suppose that the $j$-th DS is connected with $r\leq m$ origin nodes, with $r>0$, else the destination node is not connected and the lemma holds trivially for the particular $j$ and $\forall k \in \cK$. We show that the lemma holds for any $j, t, k$ and therefore it will hold for the sum over all $j\in\cJ, t\in\cT, k\in\cK$. We consider the cases where $d_{jtk}>0$ since in the case of $d_{jtk}=0$, the lemma holds trivially for the $(j, t, k)$-term. Further, we assume that at least one $a_{ik} = 1$ for the $r$ connections of DS node $j$ otherwise the lemma holds again trivially for the particular term. Given the above assumptions, without loss of generality we set all $a_{ik}$ and $d_{jtk}$ parameters to 1 for clarity of exposition.

Denote as $\cI_j$ the set of FCs that the $j$-th DS is connected, with $|\cI_j| = r$, and set $\bar{t} = T - t + 1$. From the inequality of arithmetic and geometric means we get (for fixed $j, t, k$):
\begin{equation}
\label{eq:am_gm}
\begin{aligned}
    \prod_{i\in\cI_j, \tau \geq t}(1 - x_{ij\tau}) & \leq \left(\frac{\sum_{i\in\cI_j, \tau \geq t}\left(1 - x_{ij\tau}\right)}{r\bar{t}}\right)^{r\bar{t}} \\
    & = \left(1 -\frac{\sum_{i\in\cI_j, \tau \geq t} x_{ij\tau}}{r\bar{t}}\right)^{r\bar{t}}.
\end{aligned}
\end{equation}
Let $\zeta(z) = 1 - \left(1 - \frac{z}{r\bar{t}}\right)^{r\bar{t}}$. The function $\zeta(z)$ is concave in the interval $z\in[0, 1]$ which implies that $\zeta(z) \geq \rho(r\bar{t})z$. Using this result in \eqref{eq:am_gm} we get:
\begin{equation}
\label{eq:rho_proof}
\begin{aligned}
    1 - \prod_{i\in\cI_j, \tau \geq t}(1 - x_{ij\tau}) & \geq 1 -\left(1 - \frac{\sum_{i\in\cI_j, \tau \geq t} x_{ij\tau}}{r\bar{t}}\right)^{r\bar{t}} = \zeta\left(\sum_{i\in\cI_j, \tau \geq t}x_{ij\tau}\right) \\
    & \geq  \rho(r\bar{t}) \sum_{i\in\cI_j, \tau \geq t}x_{ij\tau} \geq \rho(r\bar{t}) \min \left(1, \sum_{i\in\cI_j, \tau \geq t}x_{ij\tau} \right) \\ & \geq \rho(mT) \min \left(1, \sum_{i\in\cI_j, \tau \geq t}x_{ij\tau} \right),
\end{aligned}
\end{equation}
where the last inequality holds since $r\bar{t} \leq mT$ implies $\rho(r\bar{t}) \geq \rho(mT)$. Summing both sides of \eqref{eq:rho_proof} over all $j\in\cD, t\in\cT, k\in\cK$ proves the lemma.

\subsection{Proof of Lemma~\ref{lemma:approx}}
\label{app:proof_lemma_approx}
Since the pipage step integralizes at least one variable at a time, it is easy to see that the pipage rounding algorithm will stop in at most $I\cdot J \cdot T$ iterations. We get:
\begin{equation*}
    \texttt{opt} \leq f(\vX^{(0)}) \leq \frac{1}{\rho(mT)} g(\vX^{(0)})
    \leq \frac{1}{\rho(mT)} g(\vX^{(L)}), 
\end{equation*}
where, the first inequality is from \eqref{eq:opt_f}, the second inequality holds due to Lemma \ref{lemma:rho}, and the last one uses  the property of pipage steps $g(\vX^{(\ell)})\leq g(\vX^{(\ell+1)}),~~\forall \ell=0,\dots,L-1$.

\section{Datasets}
\label{app:data}
\textit{Network configuration:} We consider a square map with area 1200km$^2$ and a given network size, i.e., number of FCs and DSs. For each node we sample sequentially its coordinates in order to respect some distance constraints from the already created nodes. In particular, each FC needs to be at least 100km away from other FCs and each DS at least 30km away from all other nodes (FCs and DSs). In case no feasible location can be sampled, these limits are iteratively relaxed by a factor (0.99), until a feasible location is sampled.

Based on the Euclidean distance $dist_{ij}$ (in km) of all $(i, j)$ pairs, with $i\in\cI$ and $j\in\cJ$, and an average driving speed sampled from $v_{ij} \sim U[60, 80]$ km/h (we define by $U[a, b]$ the continuous uniform distribution on the interval $[a, b]$), we compute the transit time $\delta_{ij} = dist_{ij} / v_{ij}$. We assume than an FC-DS connection exists if (i) the DS is in the 50\% closest (in terms of transit time) DSs to the FC or (ii) if the FC is in the 25\% closest FCs to the DS. This distinction is important to avoid isolated nodes: if a DS is far from all FCs then no FC will select to connect to it. On the other hand, if an FC is isolated then no DS will connect to it. By creating connections based on both FC and DS proximity we ensure a more balanced and connected network. Note that with the above procedure an FC is connected to no more than 75\% of the DSs.  

\begin{figure}
\centering
\begin{subfigure}{0.46\textwidth}
    \includegraphics[width=0.99\linewidth]{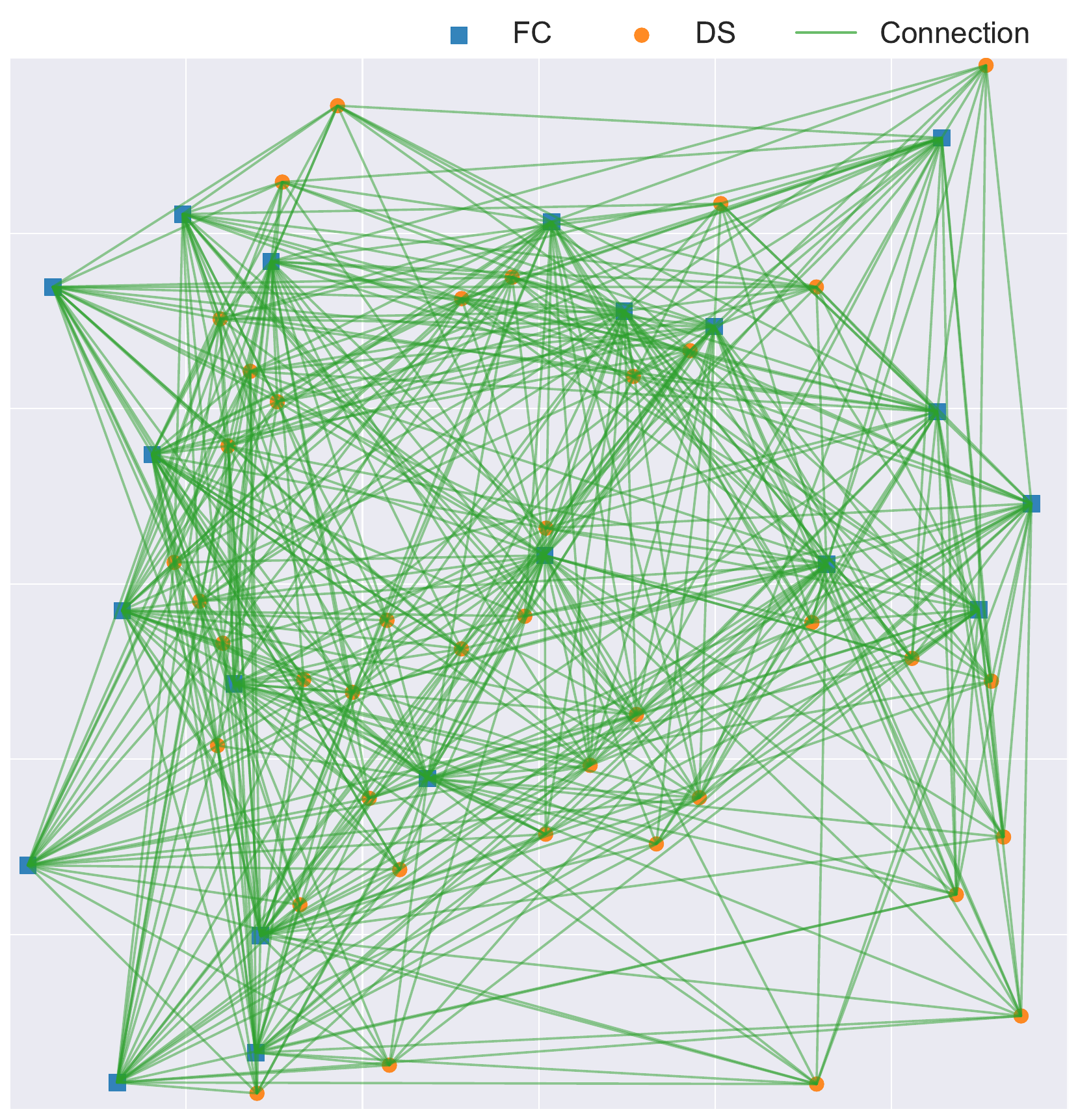}
    \caption{Example network configuration with 20 FCs and 40 DSs.}
    \label{fig:network}
\end{subfigure}
\hfill
\begin{subfigure}{0.46\textwidth}
    \includegraphics[width=0.99\linewidth]{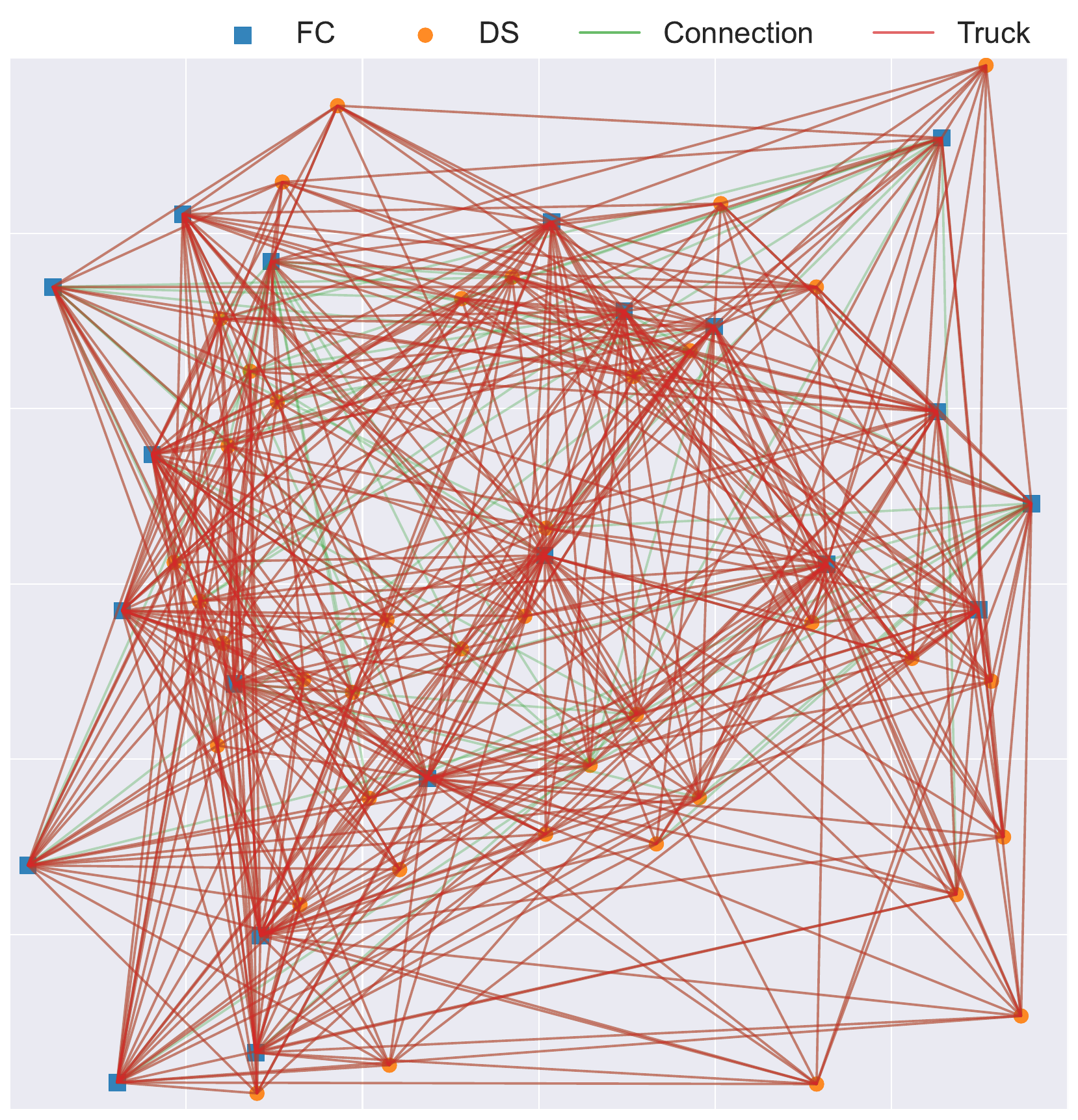}
    \caption{Active connections after optimization, i.e., connections with placed trucks.}
    \label{fig:network_solved}
\end{subfigure}
\hfill
\begin{subfigure}{0.85\textwidth}
    \includegraphics[width=0.95\textwidth]{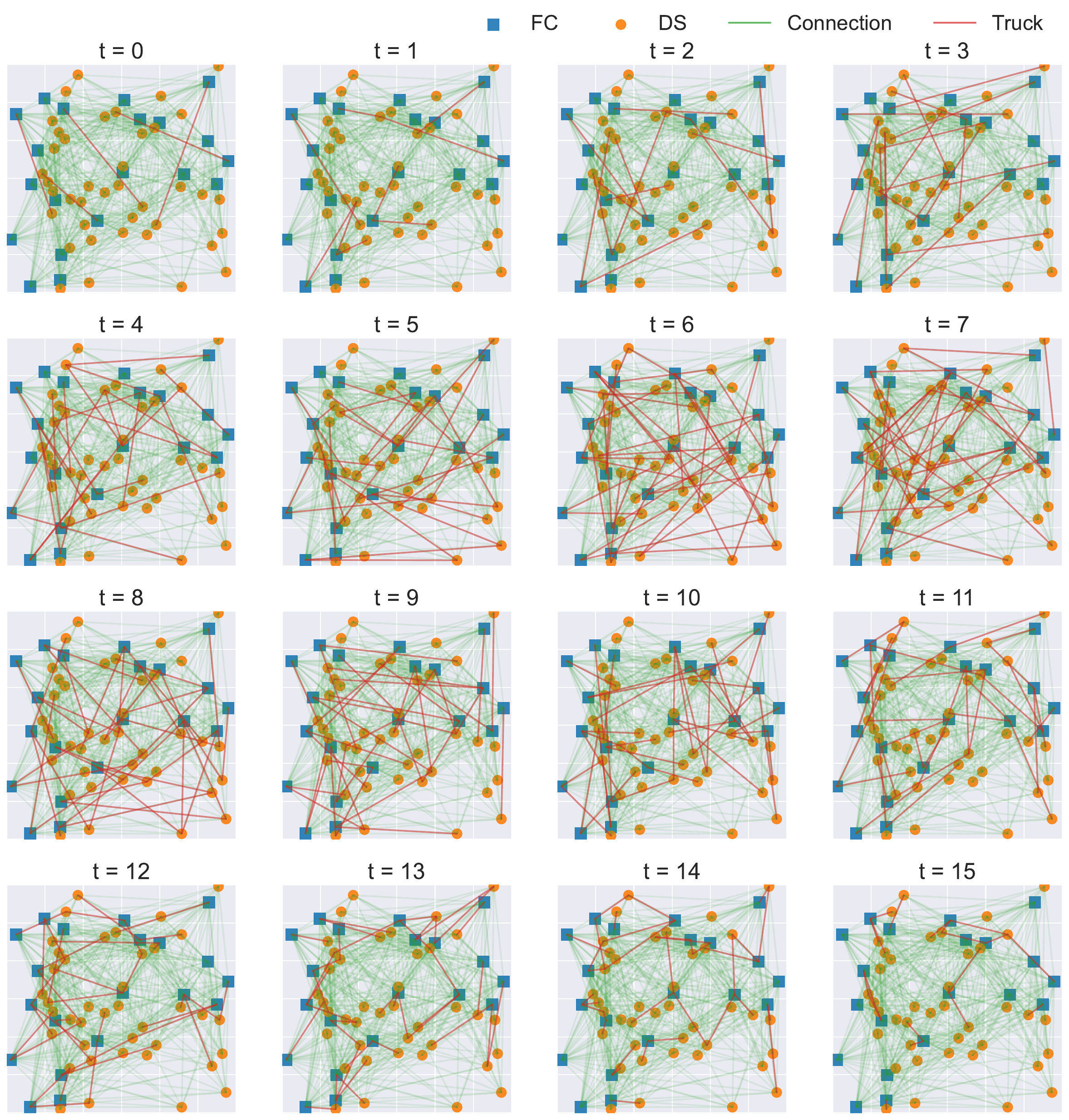}
    \caption{Truck placement at each connection per timeslot.}
    \label{fig:network_solved_slot}
\end{subfigure}
        
\caption{Network generation and last truck placement example.}
\label{fig:nets}
\end{figure}

Figure \ref{fig:network} depicts an example of such a configuration, with 20 FCs and 40 DSs. The green lines correspond to connections. Figure \ref{fig:network_solved} shows an indicative solution, with the red lines corresponding to connections that a truck was placed (at any timeslot). This solution is broken down to timeslots (with hourly granularity) in Figure~\ref{fig:network_solved_slot}. The time index over each subfigure in Figure~\ref{fig:network_solved_slot} is relative to the first timeslot that trucks were placed.

\captionsetup[sub]{font=small}

\begin{figure}[t]
\centering
  \includegraphics[width=0.7\linewidth]{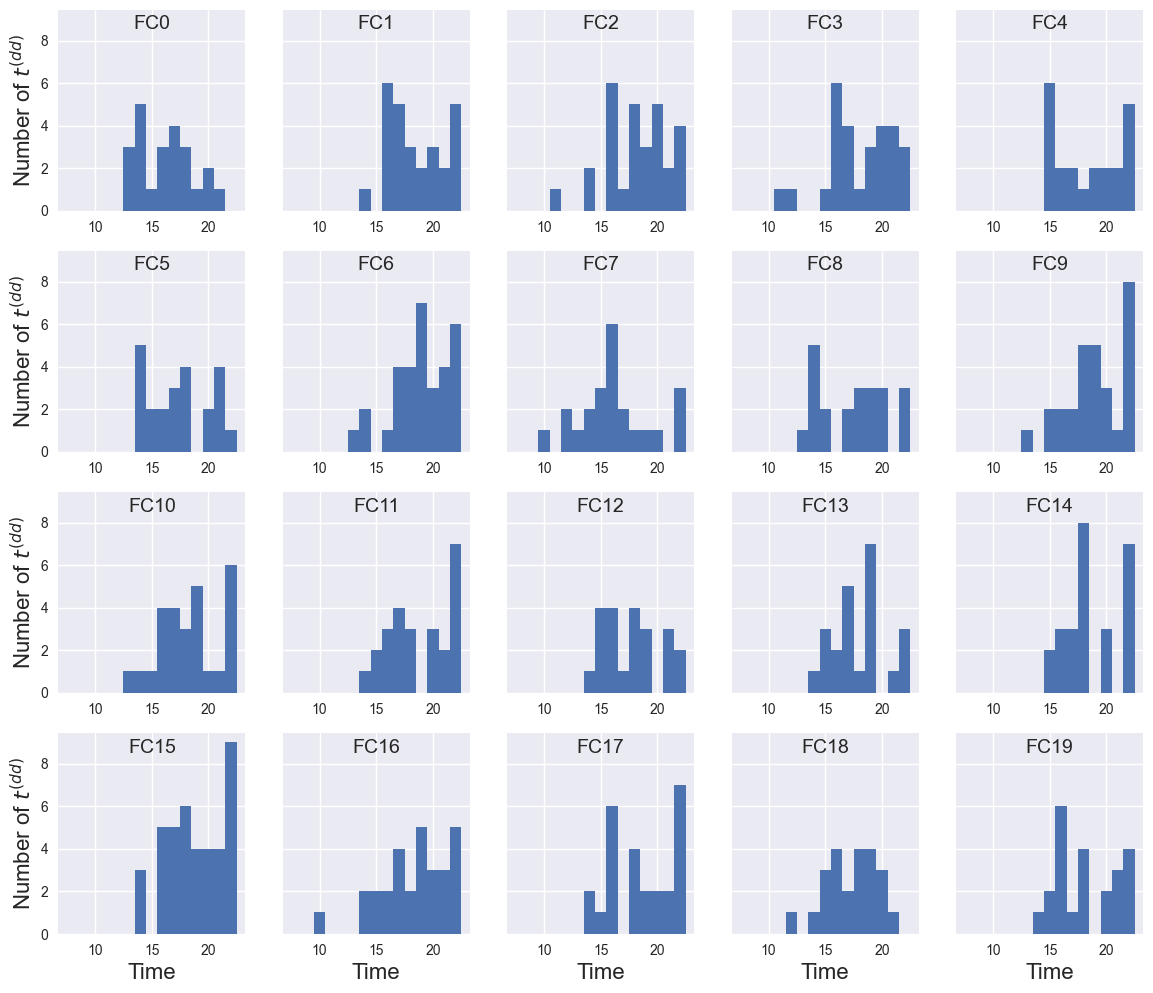}
  \caption{Histogram of the deadline departure times of all the connections of each FC.}
  \label{fig:lop}
\end{figure}

\textit{Arrival and departure deadlines:} Each DS $j\in\cJ$ has an arrival deadline sampled from $t^\text{(ad)}_j\sim U\{22, 23, \dots, 27\}$ (we define by $U\{a, \dots, b\}$ the discrete uniform distribution on the set $\{a, \dots, b\}$), i.e, from 10pm the same day till 5am the next morning. The departure deadline of each FC $i\in\cI$ towards DS $j\in\cJ$ can be computed as $t^\text{(dd)}_{ij} = \max(t^\text{(ad)}_j - \delta_{ij}, 0)$. In Figure \ref{fig:lop} we present the histogram of the deadline departure times of all the connections of each FC for the example network configuration in Figure \ref{fig:network}. 
 
\begin{figure}
\centering
  \includegraphics[width=0.7\linewidth]{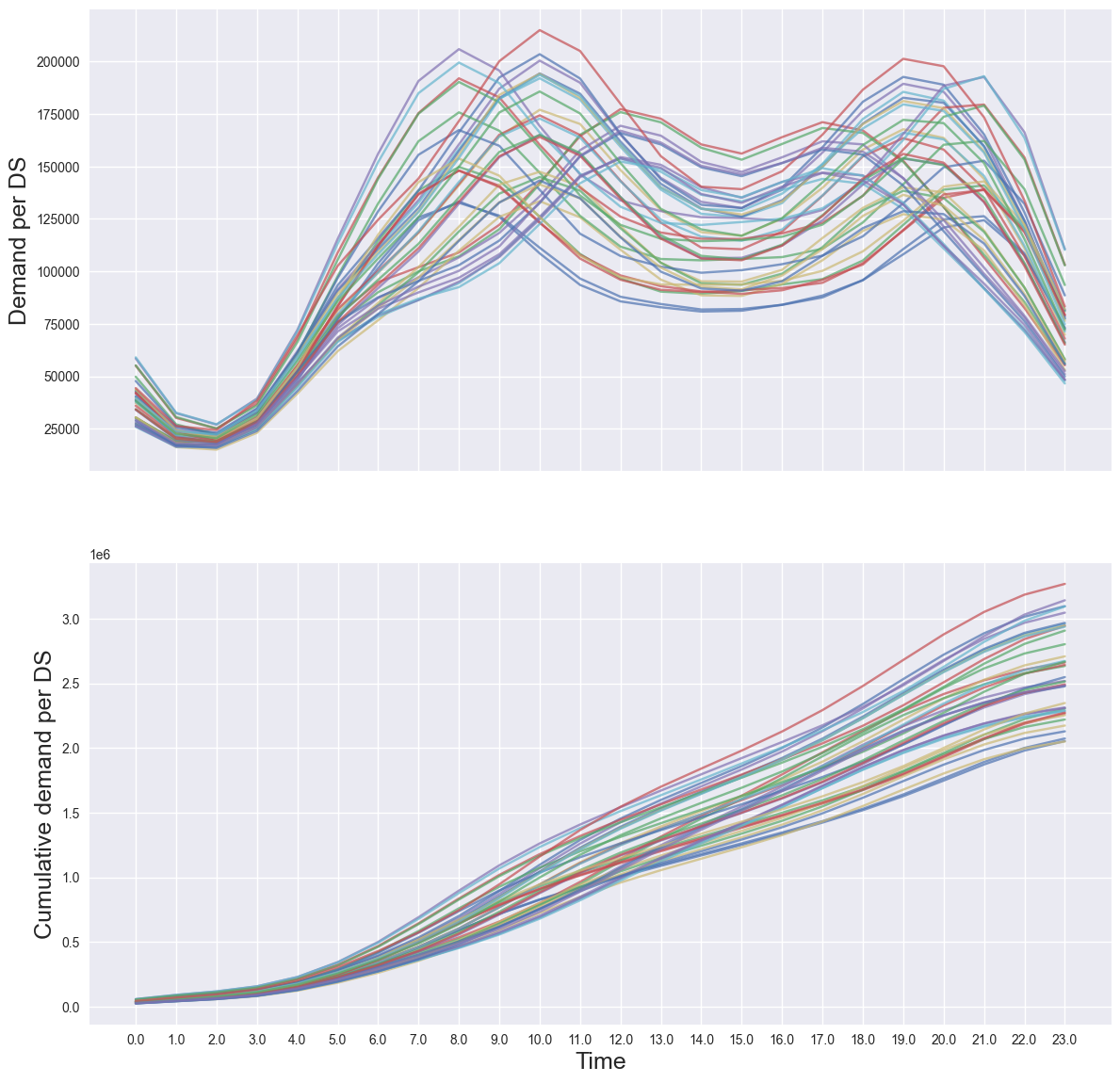}
  \caption{Total demand (over all product categories) per timeslot of each DS.}
  \label{fig:demand}
\end{figure}

\textit{Product categories and demand:} For a given set of product categories $\cK$, each category $k\in\cK$ contains a number of products drawn from a uniform distribution  $U\{100, 101, \dots, 200\}$. Each category has a mean product demand $m_k$, which models the general expected trend of the category's products, e.g., a furniture category is expected to have less orders than a book category. Each product in a category has its own expected demand profile, which is a sample drawn from the normal distribution $\cN(m_k, 0.1m_k)$.    

Each FC has only a subset of 20-25\% of all the categories available, selected at random. If a category is available in an FC, it is assumed that all the corresponding products are available. Each DS has a demand profile that models the pattern of orders during the day, e.g., one DS might usually have more requests during morning hours, while another DS afternoon or late night hours. These patterns are modeled using a mixture of Gaussian distributions, where some components are completely random and some impose specific behaviors. Finally, each DS has requests from each category with a probability sampled from $U[0.5, 1]$. If there are requests from a category, for each product of the category we sample the number of items requested based on the expected demand profile of that product. The timeslot the product is requested is sampled based on the DS demand profile.

Figure \ref{fig:demand} illustrates the demand per timeslot of all DSs. Each line corresponds to a different DS and it has a different demand profile (mixture of Gaussians) than the other DSs. Also, note the difference in the absolute number of orders. Some DSs have requests from more categories (or more popular categories) than others, which makes the total number of requests much higher.

%% file: main.bbl
\begin{thebibliography}{40}
\providecommand{\natexlab}[1]{#1}
\providecommand{\url}[1]{\texttt{#1}}
\expandafter\ifx\csname urlstyle\endcsname\relax
  \providecommand{\doi}[1]{doi: #1}\else
  \providecommand{\doi}{doi: \begingroup \urlstyle{rm}\Url}\fi

\bibitem[Ageev and Sviridenko(2004)]{ageev2004pipage}
Alexander~A Ageev and Maxim~I Sviridenko.
\newblock Pipage rounding: A new method of constructing algorithms with proven
  performance guarantee.
\newblock \emph{Journal of Combinatorial Optimization}, 8\penalty0
  (3):\penalty0 307--328, 2004.

\bibitem[Banerjee et~al.(2022)Banerjee, Erera, and
  Toriello]{doi:10.1287/trsc.2022.1125}
Dipayan Banerjee, Alan~L. Erera, and Alejandro Toriello.
\newblock Fleet sizing and service region partitioning for same-day delivery
  systems.
\newblock \emph{Transportation Science}, 56\penalty0 (5):\penalty0 1327--1347,
  2022.

\bibitem[Barnhart and Schneur(1996)]{barnhart-express96}
Cynthia Barnhart and Rina~R Schneur.
\newblock Air network design for express shipment service.
\newblock \emph{Operations Research}, 44\penalty0 (6):\penalty0 852--863, 1996.

\bibitem[Calinescu et~al.(2011)Calinescu, Chekuri, Pal, and
  Vondr{\'a}k]{calinescu2011maximizing}
Gruia Calinescu, Chandra Chekuri, Martin Pal, and Jan Vondr{\'a}k.
\newblock Maximizing a monotone submodular function subject to a matroid
  constraint.
\newblock \emph{SIAM Journal on Computing}, 40\penalty0 (6):\penalty0
  1740--1766, 2011.

\bibitem[Campbell(2009)]{hub-location-campbell}
F.~Campbell.
\newblock Hub location for time definite transportation.
\newblock \emph{Computers \& Operations Research}, 36\penalty0 (12):\penalty0
  3107--3116, 2009.

\bibitem[Carbone et~al.(2017)Carbone, Rouquet, and Roussat]{carbone2017rise}
Valentina Carbone, Aur{\'e}lien Rouquet, and Christine Roussat.
\newblock The rise of crowd logistics: a new way to co-create logistics value.
\newblock \emph{Journal of Business Logistics}, 38\penalty0 (4):\penalty0
  238--252, 2017.

\bibitem[Chekuri and Kumar(2004)]{chekuri-max-budget}
Chandra Chekuri and Amit Kumar.
\newblock Maximum coverage problem with group budget constraints and
  applications.
\newblock \emph{APPROX-RANDOM}, pages 72--83, 2004.

\bibitem[Dellaert et~al.(2019)Dellaert, Saridarq, Woensel, and
  Crainic]{dellaert2019}
Nico Dellaert, Fardin~Dashty Saridarq, Tom~Van Woensel, and Teodor~Gabriel
  Crainic.
\newblock Branch-and-price–based algorithms for the two-echelon vehicle
  routing problem with time windows.
\newblock \emph{Transportation Science}, 53\penalty0 (2):\penalty0 463--479,
  2019.

\bibitem[Deshpande and Pendem(2022)]{deshpande2022logistics}
Vinayak Deshpande and Pradeep~K Pendem.
\newblock Logistics performance, ratings, and its impact on customer purchasing
  behavior and sales in e-commerce platforms.
\newblock \emph{Manufacturing \& Service Operations Management}, 2022.

\bibitem[Feige(1998)]{feige}
Uriel Feige.
\newblock A threshold of $\ln n$ for approximating set cover.
\newblock \emph{Journal of ACM}, 45\penalty0 (4):\penalty0 634--652, 1998.

\bibitem[{FICO}(2022)]{fico-xpress}
{FICO}.
\newblock Xpress optimizer, 2022.
\newblock URL \url{https://www.fico.com/en/ products/fico-xpress-solver}.

\bibitem[Figliozzi(2020)]{figliozzi2020carbon}
Miguel~A Figliozzi.
\newblock Carbon emissions reductions in last mile and grocery deliveries
  utilizing air and ground autonomous vehicles.
\newblock \emph{Transportation Research Part D: Transport and Environment},
  85:\penalty0 102443, 2020.

\bibitem[Filmus and Ward(2012)]{filmus-local-2012}
Yuval Filmus and Justin Ward.
\newblock {The Power of Local Search: Maximum Coverage over a Matroid}.
\newblock In \emph{Proceedings of STACS}, pages 601--612, 2012.

\bibitem[Fisher(2004)]{fischer-lagrange}
Marshall~L. Fisher.
\newblock The lagrangian relaxation method for solving integer programming
  problems.
\newblock \emph{Management Science}, 50\penalty0 (12), 2004.

\bibitem[Greening et~al.(2023)Greening, Dahan, and Erera]{dahan-lead-time}
Lacy~M. Greening, Mathieu Dahan, and Alan~L. Erera.
\newblock Lead-time-constrained middle-mile consolidation network design with
  fixed origins and destinations.
\newblock \emph{Transportation Research Part B: Methodological}, 174:\penalty0
  102782, 2023.

\bibitem[{Gurobi Optimization, LLC}(2023)]{gurobi}
{Gurobi Optimization, LLC}.
\newblock {Gurobi Optimizer Reference Manual}, 2023.
\newblock URL \url{https://www.gurobi.com}.

\bibitem[Hewitt and Lehuede(2023)]{hewitt-ssndp}
Mike Hewitt and Fabien Lehuede.
\newblock New formulations for the scheduled service network design problem.
\newblock \emph{Transportation Research Part B: Methodological}, 172:\penalty0
  117--133, 2023.

\bibitem[Klapp et~al.(2018{\natexlab{a}})Klapp, Erera, and
  Toriello]{dispatch-TransScienc2018}
Mathias~A Klapp, Alan~L Erera, and Alejandro Toriello.
\newblock The one-dimensional dynamic dispatch waves problem.
\newblock \emph{Transportation Science}, 52\penalty0 (2):\penalty0 402--415,
  2018{\natexlab{a}}.

\bibitem[Klapp et~al.(2018{\natexlab{b}})Klapp, Erera, and
  Toriello]{dispatch-TransScienc2018b}
Mathias~A. Klapp, Alan~L Erera, and Alejandro Toriello.
\newblock The dynamic dispatch waves problem for same-day delivery.
\newblock \emph{European Journal of Operational Research}, 271\penalty0
  (2):\penalty0 519--534, 2018{\natexlab{b}}.

\bibitem[Klein and Steinhardt(2022)]{klein2022dynamic}
Vienna Klein and Claudius Steinhardt.
\newblock Dynamic demand management and online tour planning for same-day
  delivery.
\newblock \emph{European Journal of Operational Research}, 2022.

\bibitem[Krause and Golovin(2014)]{krause2014submodular}
Andreas Krause and Daniel Golovin.
\newblock Submodular function maximization.
\newblock \emph{Tractability}, 3:\penalty0 71--104, 2014.

\bibitem[Lara et~al.(2023)Lara, Koenemann, Nie, and
  Souza]{konemann-time-expanded}
CL~Lara, J~Koenemann, Y~Nie, and CC~de Souza.
\newblock Scalable timing-aware network design via lagrangian decomposition.
\newblock \emph{European Journal of Operational Research}, 309\penalty0
  (1):\penalty0 152--169, 2023.

\bibitem[Lee et~al.(2010)Lee, Sviridenko, and Vondr{\'a}k]{lee-mathOR2010}
John Lee, Maxim Sviridenko, and Jan Vondr{\'a}k.
\newblock Submodular maximization over multiple matroids via generalized
  exchange properties.
\newblock \emph{Mathematics of Operations Research}, 35\penalty0 (4):\penalty0
  795--806, 2010.

\bibitem[{Lee, J., V. Mirrokni, V. Nagarajan, M.
  Sviridenko}(2010)]{lee-siam-2010}
{Lee, J., V. Mirrokni, V. Nagarajan, M. Sviridenko}.
\newblock Maximizing non-monotone submodular functions under matroid and
  knapsack constraints.
\newblock \emph{SIAM J. Discrete Math.}, 23\penalty0 (4), 2010.

\bibitem[Mollenkopf et~al.(2020)Mollenkopf, Ozanne, and
  Stolze]{mollenkopf2020transformative}
Diane~A Mollenkopf, Lucie~K Ozanne, and Hannah~J Stolze.
\newblock A transformative supply chain response to covid-19.
\newblock \emph{Journal of Service Management}, 2020.

\bibitem[Nemhauser et~al.(1978)Nemhauser, Wolsey, and
  Fisher]{nemhauser1978analysis}
George~L Nemhauser, Laurence~A Wolsey, and Marshall~L Fisher.
\newblock An analysis of approximations for maximizing submodular set
  functions—i.
\newblock \emph{Mathematical programming}, 14\penalty0 (1):\penalty0 265--294,
  1978.

\bibitem[Otto et~al.(2018)Otto, Agatz, Campbell, Golden, and
  Pesch]{otto2018optimization}
Alena Otto, Niels Agatz, James Campbell, Bruce Golden, and Erwin Pesch.
\newblock Optimization approaches for civil applications of unmanned aerial
  vehicles {(UAVs)} or aerial drones: A survey.
\newblock \emph{Networks}, 72\penalty0 (4):\penalty0 411--458, 2018.

\bibitem[Polyak(1987)]{polyak-step}
Boris Polyak.
\newblock Introduction to optimization.
\newblock \emph{Translations Series in Mathematics and Engineering}, 1987.

\bibitem[Preston et~al.(2020)Preston, Nicholas, Matthews, and
  Martello]{generation-report-2020}
Felix Preston, Kukrika Nicholas, H.~Scott Matthews, and Miguel H.~Jaller
  Martello.
\newblock {GHG} footprint: Ecommerce vs brick and mortar.
\newblock \emph{Generation Investment Management LLP}, 2020.

\bibitem[Savelsbergh and Van~Woensel(2016)]{savelsb-tutorial-TS16}
Martin Savelsbergh and Tom Van~Woensel.
\newblock 50th anniversary invited article—{C}ity logistics: Challenges and
  opportunities.
\newblock \emph{Transportation Science}, 50\penalty0 (2):\penalty0 579--590,
  2016.

\bibitem[Schrijver et~al.(2003)]{schrijver2003combinatorial}
Alexander Schrijver et~al.
\newblock \emph{Combinatorial optimization: polyhedra and efficiency},
  volume~24.
\newblock Springer, 2003.

\bibitem[Shanmugam et~al.(2013)Shanmugam, Golrezaei, Dimakis, Molisch, and
  Caire]{femtocaching}
Karthikeyan Shanmugam, Negin Golrezaei, Alexandros~G. Dimakis, Andreas~F.
  Molisch, and Giuseppe Caire.
\newblock Femtocaching: Wireless content delivery through distributed caching
  helpers.
\newblock \emph{IEEE Transactions on Information Theory}, 59\penalty0
  (12):\penalty0 8402--8413, 2013.

\bibitem[Sluijk et~al.(2023{\natexlab{a}})Sluijk, Florio, Kinable, Dellaert,
  and Van~Woensel]{Sluijk2023a}
Natasja Sluijk, Alexandre~M Florio, Joris Kinable, Nico Dellaert, and Tom
  Van~Woensel.
\newblock A chance-constrained two-echelon vehicle routing problem with
  stochastic demands.
\newblock \emph{Transportation Science}, 57\penalty0 (1):\penalty0 252--272,
  2023{\natexlab{a}}.

\bibitem[Sluijk et~al.(2023{\natexlab{b}})Sluijk, Florio, Kinable, Dellaert,
  and Van~Woensel]{Sluijk2023b}
Natasja Sluijk, Alexandre~M Florio, Joris Kinable, Nico Dellaert, and Tom
  Van~Woensel.
\newblock Two-echelon vehicle routing problems: A literature review.
\newblock \emph{European Journal of Operational Research}, 304\penalty0
  (3):\penalty0 865--886, 2023{\natexlab{b}}.

\bibitem[Stroh et~al.(2022)Stroh, Erera, and
  Toriello]{doi:10.1287/mnsc.2021.4041}
Alexander~M. Stroh, Alan~L. Erera, and Alejandro Toriello.
\newblock Tactical design of same-day delivery systems.
\newblock \emph{Management Science}, 68\penalty0 (5):\penalty0 3444--3463,
  2022.

\bibitem[Voccia et~al.(2019)Voccia, Campbell, and
  Thomas]{ThomasBarrettW2019TSDP}
Stacy~A Voccia, Ann~Melissa Campbell, and Barrett~W Thomas.
\newblock The same-day delivery problem for online purchases.
\newblock \emph{Transportation Science}, 53\penalty0 (1):\penalty0 167--184,
  2019.
\newblock ISSN 0041-1655.

\bibitem[Wu et~al.(2023)Wu, Herszterg, Savelsbergh, and Huang]{wu-snd-ts23}
Haotian Wu, Ian Herszterg, Martin Savelsbergh, and Yixiao Huang.
\newblock Service network design for same-day delivery with hub capacity
  constraints.
\newblock \emph{Transportation Science}, 57\penalty0 (1):\penalty0 273--287,
  2023.

\bibitem[Yaman et~al.(2012)Yaman, Karasan, and
  Kara]{doi:10.1287/opre.1120.1065}
Hande Yaman, Oya~Ekin Karasan, and Bahar~Y. Kara.
\newblock Release time scheduling and hub location for next-day delivery.
\newblock \emph{Operations Research}, 60\penalty0 (4):\penalty0 906--917, 2012.

\bibitem[Yildiz et~al.(2021)Yildiz, Yaman, and Karasan]{hub-location-Transp21}
Baris Yildiz, Hande Yaman, and Oya~Ekin Karasan.
\newblock Hub location, routing, and route dimensioning: Strategic and tactical
  intermodal transportation hub network design.
\newblock \emph{Transportation Science}, 55\penalty0 (6):\penalty0 1351--1369,
  2021.

\bibitem[Yıldız and Savelsbergh(2022)]{yildiz-package-express}
Barış Yıldız and Martin Savelsbergh.
\newblock Optimizing package express operations in china.
\newblock \emph{European Journal of Operational Research}, 300\penalty0
  (1):\penalty0 320--335, 2022.

\end{thebibliography}
